\documentclass[pra,reprint,superscriptaddress]{revtex4-1}
\usepackage{xcolor}
\usepackage{graphicx}% Include figure files
\usepackage{dcolumn}% Align table columns on decimal point
\usepackage{bm}% bold math
\usepackage[braket, qm]{qcircuit}
\usepackage{amsmath}
\usepackage{multirow}
\pdfmapfile{+sansmathaccent.map}
\definecolor{forestgreen}{rgb}{0.0, 0.27, 0.13}

\usepackage[T1]{fontenc}
\usepackage[utf8]{inputenc}
\begin{document}

    %\preprint{APS/123-QED}
    
    \title{Modeling of Measurement-based Quantum Network Coding on IBM Q Experience Devices}
    \author{Poramet Pathumsoot}
    \affiliation{Department of Physics, Faculty of Science, Mahidol University, Bangkok 10400, Thailand}
    \author{Takaaki Matsuo}
    \affiliation{ Keio University Shonan Fujisawa Campus, 5322 Endo, Fujisawa, Kanagawa 252-0882, Japan}
    \author{Takahiko Satoh}
    \affiliation{ Keio University Shonan Fujisawa Campus, 5322 Endo, Fujisawa, Kanagawa 252-0882, Japan}
    \author{Michal Hajdu\v{s}ek}
    \affiliation{ Keio University Shonan Fujisawa Campus, 5322 Endo, Fujisawa, Kanagawa 252-0882, Japan}
    \author{Sujin Suwanna}
    \affiliation{Department of Physics, Faculty of Science, Mahidol University, Bangkok 10400, Thailand}
    \author{Rodney Van Meter}
    \email{rdv@sfc.wide.ad.jp}
    \affiliation{ Keio University Shonan Fujisawa Campus, 5322 Endo, Fujisawa, Kanagawa 252-0882, Japan}
\date{\today}

\begin{abstract}
    Quantum network coding has been proposed to improve resource utilization to support distributed computation but has not yet been put in to practice. We investigate a particular implementation of quantum network coding using measurement-based quantum computation on IBM Q processors. We compare the performance of quantum network coding with entanglement swapping and entanglement distribution via linear cluster states. These protocols outperform quantum network coding in terms of the final Bell pair fidelities but are unsuitable for optimal resource utilization in complex networks with contention present. We demonstrate the suitability of noisy intermediate-scale quantum (NISQ) devices such as IBM Q for the study of quantum networks. We also identify the factors that limit the performance of quantum network coding on these processors and provide estimates or error rates required to boost the final Bell pair fidelities to a point where they can be used for generation of genuinely random cryptographic keys among other useful tasks. Surprisingly, the required error rates are only around a factor of 2 smaller than the current status and we expect they will be achieved in the near future.
\end{abstract}
    
\pacs{Valid PACS appear here}
\maketitle
    
\section{\label{sec:Introduction}Introduction}
    Quantum communication \cite{Wehner18:eaam9288, kimble08:_quant_internet, muralidharan2016optimal, van-meter14:_quantum_networking, gisin2007quantum} is an exciting field of study encompassing numerous applications such as quantum cryptography \cite{bennett:bb84,ekert1991qcb,elliott:qkd-net,gisin2002quantum}, distributed quantum computing \cite{buhrman03:_dist_qc, van2016path} and delegated quantum computation \cite{broadbent2009universal, hajduvsek2015device,morimae2013blind,hayashi2018self}. Transmission of quantum information over long distances presents a significant technical challenge due to loss in optical channels and the sensitivity of quantum states to the environment. In order to overcome this hurdle various quantum repeater schemes were proposed \cite{briegel1998quantum,DurQuantumRepeater, PirkerLong-Range, PhysRevA.79.032325, PhysRevLett.104.180503}. These methods rely either on quantum error correction \cite{devitt13:rpp-qec} or purification \cite{dur2003mep} and combine management of loss at the link layer \cite{dahlberg2019link, matsuo2019quantum, jones16:comm-pro, humphreys2018deterministic, krutyanskiy2019light} with entanglement swapping.
        
    Experimental demonstration of a memory-based quantum repeater was achieved in \cite{yuan2008experimental} while memory-free all-photonic quantum repeaters were implemented very recently in \cite{hasegawa2019experimental,li2019experimental}. These proof-of-principle experiments are significant steps towards implementation of real quantum networks but are still incapable of being scaled up to even modestly-sized networks \cite{hensen2015loophole,rozpkedek2019near,Kumar_2019}.
        
    \begin{figure}
        \centering
        \includegraphics[width=\columnwidth]{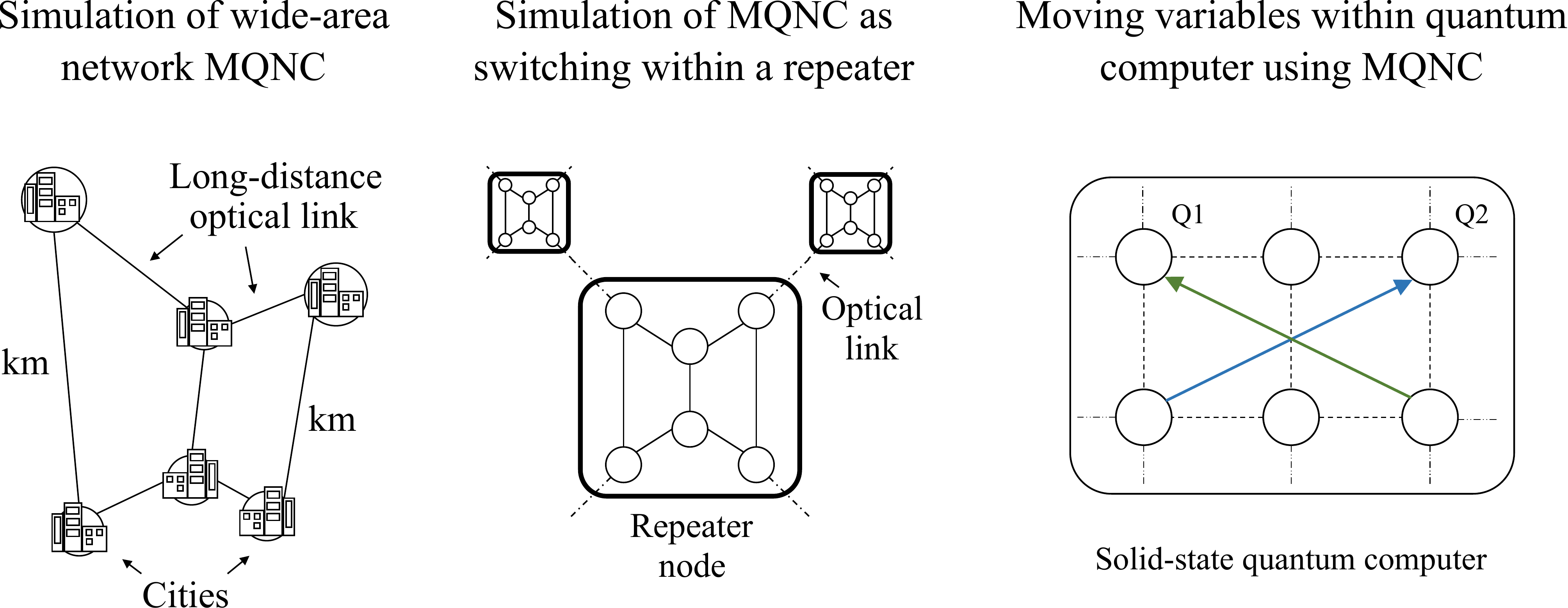}
        \caption{\label{fig:figure1_intro} The core idea behind MQNC has applications in a number of real near-future scenarios. On the left, long-distance quantum network with the outlined topology can use MQNC to establish direct links between distant cities. MQNC can also be used as a switching protocol inside repeater nodes as shown in the middle. MQNC may also be used to move the states of quantum registers in a quantum computer.}
    \end{figure}
        
    We take a step toward real-world use via a proof-of-concept implementation on a real superconducting quantum computer, IBM Q Experience device. This allows us to assess the practicality of a number of quantum protocols such as entanglement swapping \cite{zukowski1993event,EntanglementSwappingOriginal}, measurement on linear cluster states \cite{MBQC} and measurement-based quantum network coding (MQNC) \cite{MQNC} as tools for distribution of entangled qubit pairs. The former two protocols are suitable for cases when there is no contention over network resources. MQNC on the other hand is designed to prevent bottlenecks by addressing contention in the network. The motivation behind this is threefold, as shown in Fig.~\ref{fig:figure1_intro}. First, without any actual implementations of long-distance quantum networks we  are able simulate a network with quantum network coding (QNC) over long-distance links on a real device. Secondly, we simulate use of QNC inside a network node acting as a router or a switch. Lastly, this implementation points toward use of graph states as resources for moving qubits inside the physical device.
    
    \begin{figure*}
        \centering
        \includegraphics[width=\textwidth]{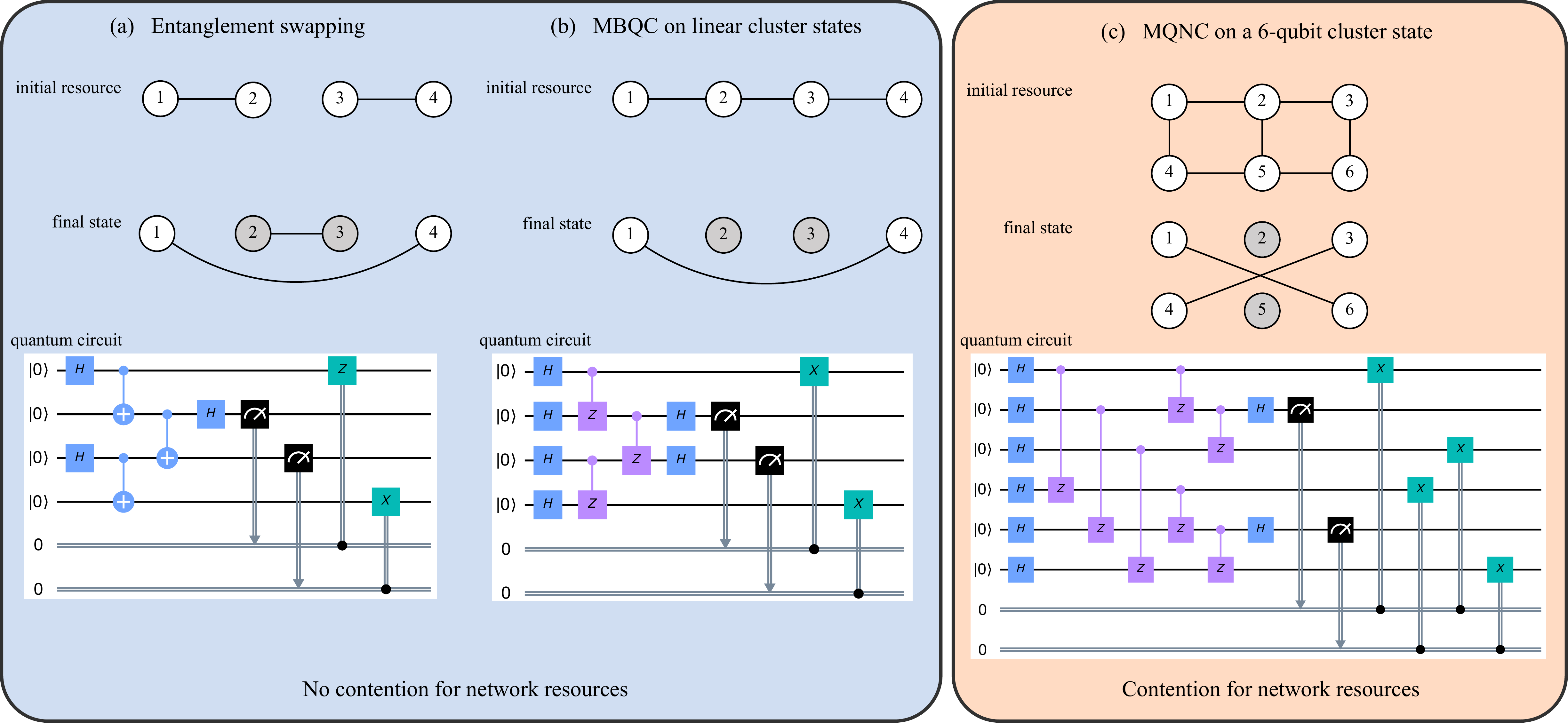}
        \caption{\label{fig:figure2_background} Three protocols for entanglement distribution that we implement. (a) Entanglement swapping uses Bell measurements to entangle previously unentagled nodes. (b) MBQC on a linear cluster state first prepares an entangled graph state and then proceeds via single-qubit measurements. Both entanglement swapping and MBQC on linear cluster states are capable of creating only a single maximally entangled pair between spatially the far nodes. When used in complex networks these protocols are not able to deal with contention for network resources. (c) MQNC on a 6-qubit cluster state. This is protocol is the latter half of the full MQNC protocol shown in Fig.~\ref{fig:figure5_mqnc}. Unlike entanglement swapping and MQBC on linear cluster states, it is capable of creating two maximally entangled pairs and can be used to deal with contention for network resources.}
    \end{figure*} 
        
    We use the IBM Q 20 Tokyo and Poughkeepsie devices to implement and compare entanglement swapping with measurement-based entanglement distribution on linear cluster states. We implement the MQNC protocol only on the IBM Q Tokyo device due to topological constraints of the Poughkeepsie device. The entanglement swapping and measurement-based entanglement distribution on linear cluster states protocols perform better in terms of the final fidelities compared to MQNC. This is not completely surprising for a number of reasons. First, IBM Q Tokyo is overall a more noisy device compared to Poughkeepsie and secondly, MQNC requires a larger number of quantum 2-qubit gates compared to the other two protocols. To study the performance of MQNC under realistic conditions we introduce a noise model and show that it describes the obtained experimental data points well. This allows us to extrapolate the error rates required for the real physical devices to implement MQNC that results in two Bell pairs with fidelities high enough to violate the CHSH inequality \cite{CHSHOriginalPaper}. We show that the error rates must decrease by at least a factor of 2 in order for this to be achieved.
    
    Our manuscript is organized as follows. In Section \ref{sec:Background}, we briefly introduce classical and quantum network coding, measurement-based quantum computing and how it can be used for long-distance communication using linear cluster states, and finally measurement-based quantum network coding. We present experimental results for contention-free communication using entanglement swapping and linear cluster states in Section \ref{sec:LinearCommunication} then implement and analyze MQNC that can handle contention in Section \ref{sec:QNConIBMQ}. Finally, we conclude the discussion in Section \ref{sec:Discussion}.

\section{\label{sec:Background}Background}

    Quantum computation, whether conducted within a monolithic quantum computer, within a quantum multicomputer \cite{van-meter07:_distr_arith_jetc} or across a network, requires us to execute gates between qubits initially held some distance apart. Within a single computer, this can be done by moving qubits by shuttling ions or using SWAP gates to bring the qubits into proximity \cite{1903.10963}. Once the qubits are brought together, two-qubit gates can be executed directly.
        
    Alternatively, we can build distributed quantum states (e.g., Bell states) that span the distance, and use those distributed states either to teleport data~\cite{bennett:teleportation} or to execute gates remotely~\cite{eisert2000oli,gottsman99:universal_teleport}.  If the qubits are more than one site apart, we can build entanglement spanning that distance via entanglement swapping~\cite{zukowski1993event,EntanglementSwappingOriginal}. This can also be achieved via linear graph states and measurement-based computation~\cite{MBQC,MBQC-2}.
        
    When the system size exceeds the capacity of a single computer, we can couple together multiple computers over optical links.  While the ideal is to transfer the state of a qubit to a photon and send it from one computer to the other, optical conversion and channel losses make that impossible. Thus, we use entanglement swapping and either purification or quantum error correction and build \emph{quantum repeaters}~\cite{briegel1998quantum,DurQuantumRepeater}.
        
    All of these methods can be used as appropriate, when the needed resources are otherwise idle.  However, when multiple operations need to happen across a topologically complex structure, \emph{contention} for resources can lead to \emph{congestion}, and force us to either alternate uses (multiplexing)~\cite{aparicio11:repeater-muxing} or build graph states that support quantum network coding.

    \subsection{\label{sec:EntanglementSwappingBackground}Entanglement swapping}
            
    With the constraint of device topology, swapping of the quantum state of two connected qubits $i$ and $j$, $\text{SWAP}|\psi\rangle_i|\phi\rangle_j=|\phi\rangle_i|\psi\rangle_j$, is often required in order to proceed further with the computation.
        
    Sometimes it is not desirable to propagate the state of a quantum register to a distant node via successive application of $\text{SWAP}$. In this case, one can use entanglement swapping \cite{zukowski1993event,EntanglementSwappingOriginal} to establish maximal entanglement between two distant qubits. Due to its fundamental role in long-distance quantum communication and quantum computation we give a brief overview of this important primitive.
            
    Consider four qubits as pictured in Fig.~\ref{fig:figure2_background}(a) where qubits $\{1,2\}$ form one Bell pair $|\Phi^+\rangle_{12}=(|00\rangle+|11\rangle)/\sqrt{2}$ while $\{3,4\}$ form another pair $|\Phi^+\rangle_{34}$. Qubits $\{1,4\}$ can be entangled by measuring $\{2,3\}$ in the Bell basis. This measurement can be implemented by applying controlled-$X$ gate, $CX_{ij}=|0\rangle\langle0|_i\otimes I_j + |1\rangle\langle1|_i\otimes X_j$, with qubit $2$ as the control and qubit $3$ as the target followed by a Hadamard, $H=|+\rangle\langle0| + |-\rangle\langle1|$, on qubit $2$. Here $|\pm\rangle=(|0\rangle + |1\rangle)/\sqrt{2}$ are the eigenstates of Pauli $X$ operator. Finally, both qubits are measured in the computational basis and the measurement outcomes are used to apply conditional byproduct transformations resulting in a Bell pair $|\Phi^+\rangle_{14}$.

    \subsection{\label{LinearClusterForLongDisCommu}Linear Cluster State for Long-Distance Communication}
        
    Measurement-based quantum computing (MBQC) \cite{MBQC,MBQC-2} is a model of universal quantum computation that uses sequential single-qubit measurements on an initial highly entangled resource state \cite{briegel2001persistent,hajduvsek2010entanglement,hajduvsek2013direct}, 
    \begin{eqnarray}\label{eq:ClusterState}
        |G\rangle = \prod_{(a, b) \in E} CZ_{ab}|+\rangle^{\otimes n}.
    \end{eqnarray}
    Here $CZ_{ab}=|0\rangle\langle0|_a\otimes I_b + |1\rangle\langle1|_a\otimes Z_b$ is the controlled-$Z$ gate acting on qubits $a$ and $b$. The entangling $CZ$ gates are applied between qubits according to the edge set $E$ of the underlying graph $G=(V,E)$ describing the topology of the network. If the underlying graph has a regular topology such as a linear chain or a 2D lattice the resource state for MBQC is usually referred to as a cluster state.
     
    MBQC on a graph state can be used as an alternative to entanglement swapping. By performing Pauli measurements and byproduct operations on the remaining qubits, which act as nodes in the network, we can transform the topology of the graph state. In particular, consider a $N$-qubit linear cluster state with open boundaries. By measuring qubits $1,\dots,N-2$ in the $X$ basis we can transform the original linear cluster state and establish a maximally entangled pair of qubits $0$ and $N-1$ \cite{MBQC}.
            
    This is pictured in Fig.~\ref{fig:figure2_background}(b). Qubits $\{1,2,3,4\}$ are entangled via $CZ$ gates resulting in a 4-qubit linear cluster state. Qubits $\{2,3\}$ are then measured in $X$ basis and conditional byproduct operators are applied on qubits $\{1,4\}$ to establish a maximally entangled state $|G_2\rangle_{14}=(|0+\rangle+|1-\rangle)/\sqrt{2}$.

    \subsection{\label{sec:MQNC}Measurement-based Quantum Network Coding}
        
    Practical quantum networks require a complex network topology interconnecting many sites. Efficiently transmitting information across a complex network requires routing (selecting a path through the network) \cite{van-meter:qDijkstra, caleffi17:routing, schoute15:_shortcut_thesis, Pant2019, gyongyosi2017entanglement,Behera2019} and management of the available resources when multiple users are concurrently requesting use of the network \cite{aparicio11:repeater-muxing}. Classical and quantum networks exhibit similar problems; for example, congestion arising at a bottleneck of the network \cite{Jacobson:1988:CAC:52324.52356}.

	\begin{figure}
		\includegraphics[width=\columnwidth]{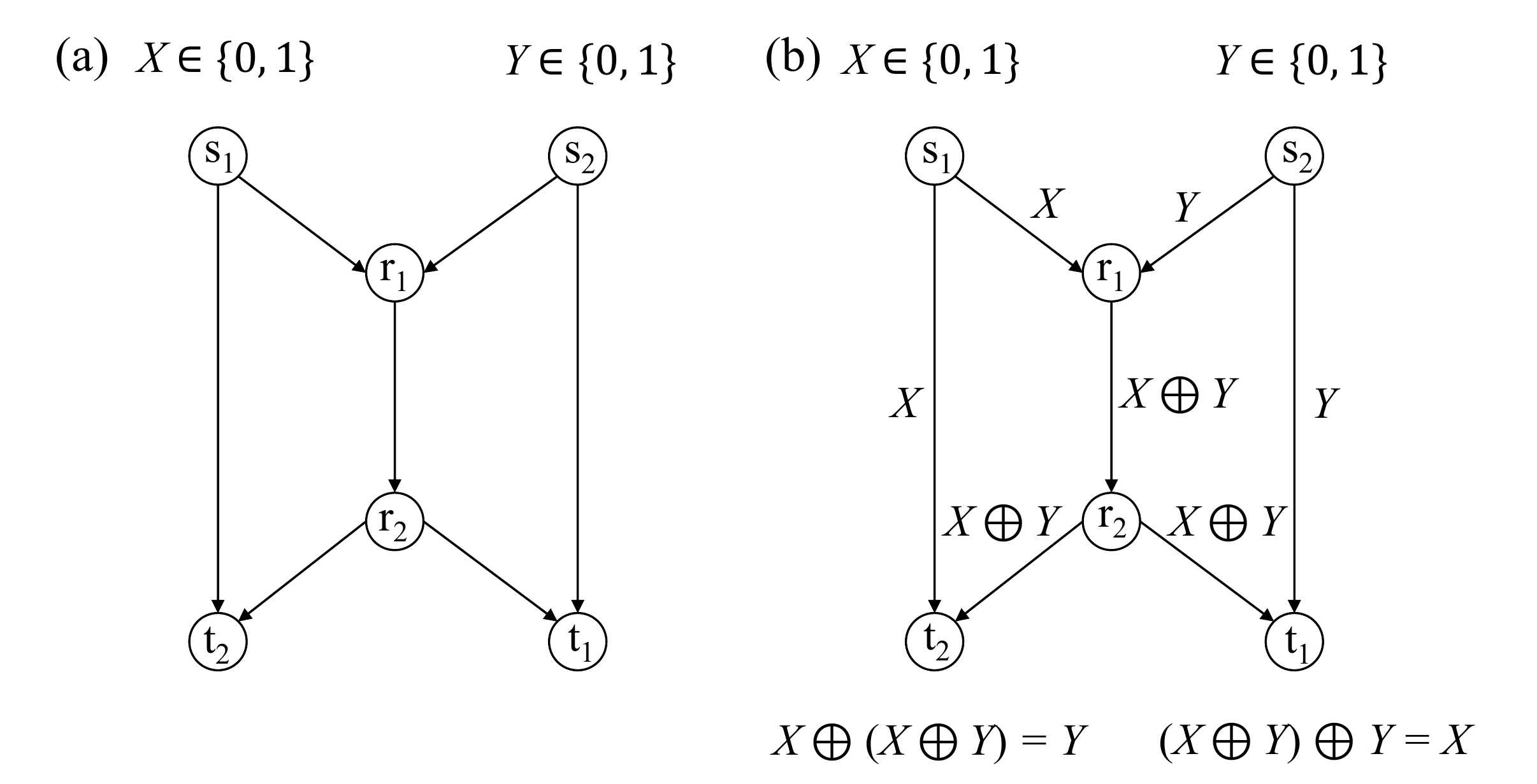}
		\caption{\label{fig:figure3_coding} The left figure shows the butterfly network topology where $s_{1}$ and $s_{2}$ each want to send a bit to $t_{1}$ and $t_{2} $ respectively. Right figure shows the classical network coding procedure using XOR operation to encode and decode the messages.}
	\end{figure}
    
    \begin{figure*}
		\centering
		\includegraphics[width=0.9\textwidth]{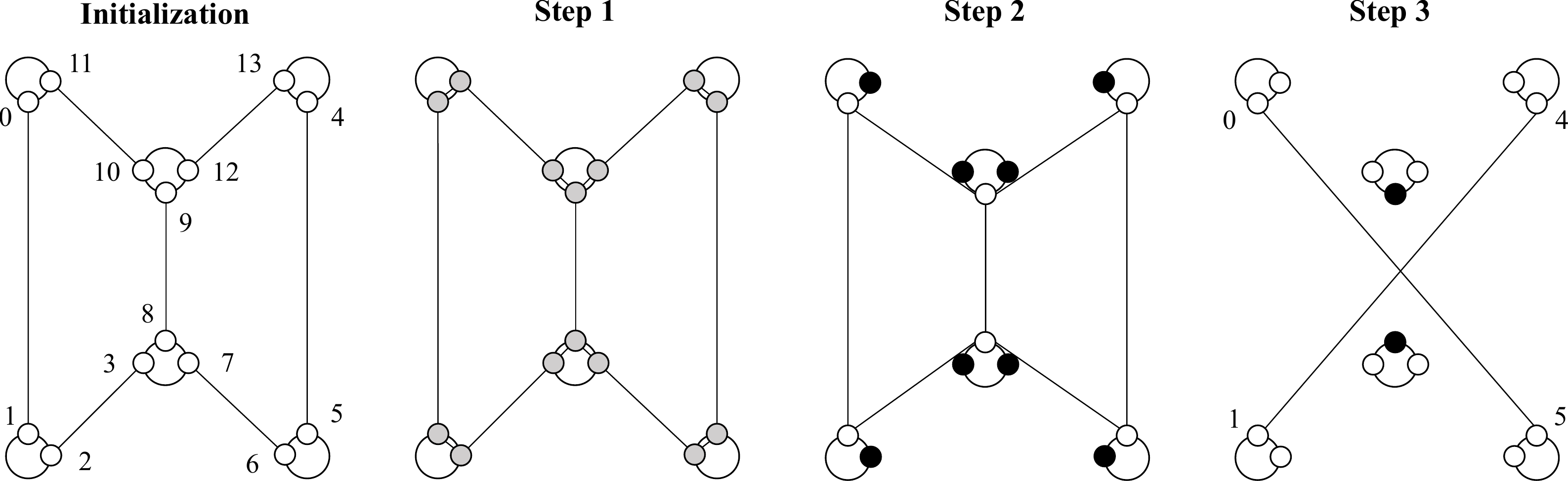}
		\caption{\label{fig:figure5_mqnc} MQNC encoding procedure. Large circles represent network nodes while small circles represent qubits. The network is initialized in a tensor product of 7 maximally entangled pairs of qubits. The qubits within each node are then entangled by $CZ$ gates as shown in Step 1. This produces a 14-qubit graph state spanning the whole network. In Step 2, the black qubits are measured in $Y$ basis. This removes them from the network and produces a cluster state on the remaining  6 qubits. Finally, the middle black qubits are measured in the $X$ basis as shown in Step 3 resulting in two 2-qubit cluster states.}
	\end{figure*}
       
    One solution to contention for access to a channel is network coding \cite{ahlswede2000nif}, best demonstrated on a butterfly network pictured in Fig.~\ref{fig:figure3_coding}. Consider classical senders $s_{1}$ and $s_{2}$ simultaneously want to send messages $X$ and $Y$ respectively, assuming that $X$ and $Y$ are both one bit of data, to their corresponding target receiver, $t_{1}$, $t_{2}$, across a bottleneck $r_{1}$ and $r_{2}$. One trivial solution is for senders to alternate use of the channel. One sender waits until the other sender successfully sends his message, takes its turn, then relinquishes the channel. Known as time division multiplexing (TDM), this simple method may under-utilize resources in a complex real-world network, forcing memories to wait and some channels to idle. Network coding, in contrast, can complete the transmission of two messages in one cycle, by encoding two incoming messages using an XOR operation at node $r_1$, then sending the encoded message to both target nodes. The remaining task is to decode the message using another XOR operation and the one message received directly from another sender \cite{ahlswede2000nif}.
       
    Similarly, in quantum communication, QNC can be used to overcome the topological limits of the network \cite{hayashi2006qnc, iwama2006qnc, leung:1840528, satoh:PhysRevA.93.032302, QNC, akibue:7556338}. Quantum network coding is primarily aimed at resource efficiency. Classical network coding has a broad range of applicability, but QNC shows an advantage over TDM in a narrower set of cases. However, it also allows us to defer routing decisions and combines communication with computation \cite{akibue:7556338, MBQC}.
        
    Inspired by QNC and measurement-based quantum computing, Matsuo \emph{et al.} introduced  measurement-based quantum network coding (MQNC) \cite{MQNC}.  Instead of relying on a Bell-pair based approach, MQNC builds a graph or cluster state and performs network coding on top of this single entangled state.
            
    MQNC aims to create two crossing-over Bell pairs as shown in Fig.~\ref{fig:figure5_mqnc} by assuming an initial shared resource. The first step is merely entangling qubits. The second step removes qubits via $Y$ measurements, consequently creating a link between neighboring qubits, resulting in a 6-qubit cluster state. The final step is to remove qubits at the bottleneck of the network via $X$ measurement resulting in 2 Bell pairs. The randomness of $X$ and $Y$ measurements introduces byproduct operations that must be tracked in order for the protocol to produce the desired quantum state.

\section{\label{sec:LinearCommunication}Linear Communication on the IBM Q Experience}

    The experiments in this paper were performed using the pure state QASM simulator, providing an ideal result, and real IBM Q Experience devices. Each trial consisted of 8192 shots. The circuit was optimized and the variable qubits mapped to the physical qubits by the Qiskit transpiler. The IBM Q Experience devices are superconducting quantum computers that are available for use across the Internet via a web-based interface or programs in Python using Qiskit libraries \cite{Qiskit}. The devices used in this paper are IBM Q 20 Tokyo and IBM Q Poughkeepsie, each having 20 qubits as shown in Fig. \ref{fig:figure5_topology}.
    
    We evaluate the performance of the real IBM Q Experience devices by computing the fidelity,
    \begin{equation}
        F(\rho, \sigma) = \left[ \text{Tr} \left\{ \sqrt{\sqrt{\rho} \sigma \sqrt{\rho}}  \right\} \right]^2,
    \end{equation}
    between the final state $\rho$ and the expected state $\sigma$. In most cases that we consider the expected state is pure, $\sigma=|\psi\rangle\langle\psi|$, and the expression for fidelity reduces to $F(\rho, |\psi\rangle) = \langle\psi| \rho | \psi\rangle$.
        
    Performing the 6-qubit cluster state model on real devices requires us to be concerned about fidelity loss along the process. To determine the expected fidelity as we add each qubit to the cluster state, we used a linear cluster state on the connected qubits of real devices. Since the devices do not support feed-forward operation, after executing a quantum circuit, data filtering is needed in order to obtain a feed-forward equivalent result, i.e., post-selection. We are assuming that there is no contention for resources. For example, when implementing quantum circuit in Fig.~\ref{fig:figure2_background}(b),  measurement outcome `1' is post-selected on qubits $\{2,3\}$.
        
    Either moving qubits via SWAP gates or executing a remote gate via a nested sequence that utilizes intermediate qubits is necessary on the IBM today. There are several options for compiling efficient circuits. Nishio \textit{et al.} \cite{1903.10963} examined the tradeoffs between various options, including error-aware compilation.

    \subsection{\label{sec:Entanglement_Swapping}Entanglement Swapping}
		
    Consider a simple model of entanglement swapping. The objective is to create a Bell pair between two distant qubits; qubit 0 will entangle with qubit 11 at the end of the operation (see selected qubit $\{0,5,6,11\}$ from Fig. \ref{fig:figure5_topology}(a)). We also conducted an experiment using qubits $\{1,5,6,10\}$. Qubits 1 and 10 will be entangled at the end of the operation. Using one trial with post-selection where the measurement results of qubits $\{5,6\}$ are `0', we performed state-tomography to reconstruct the density matrix $\rho$ from the post-selected result. We found state fidelity $F(\rho,|\Phi^+\rangle)\approx0.76$ for qubits $\{0,11\}$ and $F(\rho,|\Phi^+\rangle)\approx0.66$ for qubits $\{1,10\}$.

    \subsection{Linear cluster state}
    
    \begin{figure}
    	\centering
    	\includegraphics[width=\columnwidth]{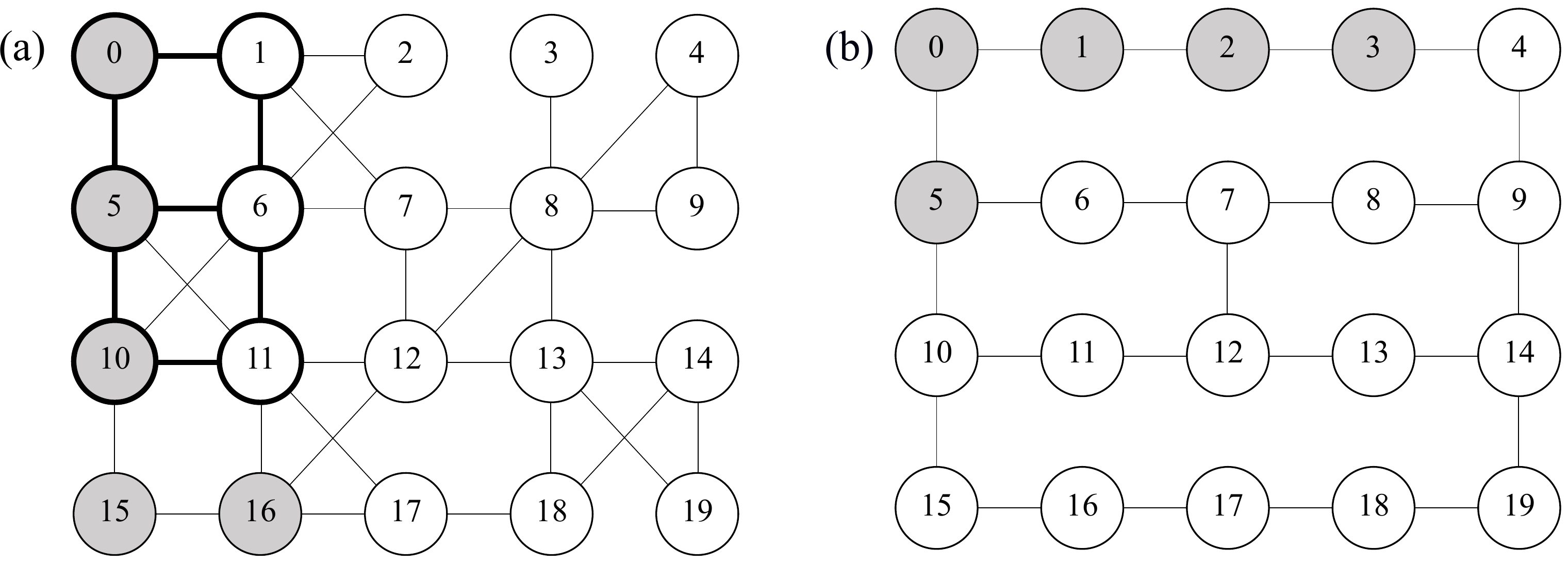}
    	\caption{\label{fig:figure5_topology} IBM Q devices topology. Gray qubits indicate qubits chosen to perform state tomography of linear cluster state. (a) IBM Q 20 Tokyo starts from qubit 0 to 16 and (b) IBM Q Poughkeepsie starts from qubit 5 to qubit 3. Heavy lines indicate qubits used to create the 6-qubit cluster state on Tokyo for both state tomography and the correlation matrix.}
    \end{figure}

	\begin{figure*}
		\centering
		\includegraphics[width=\textwidth]{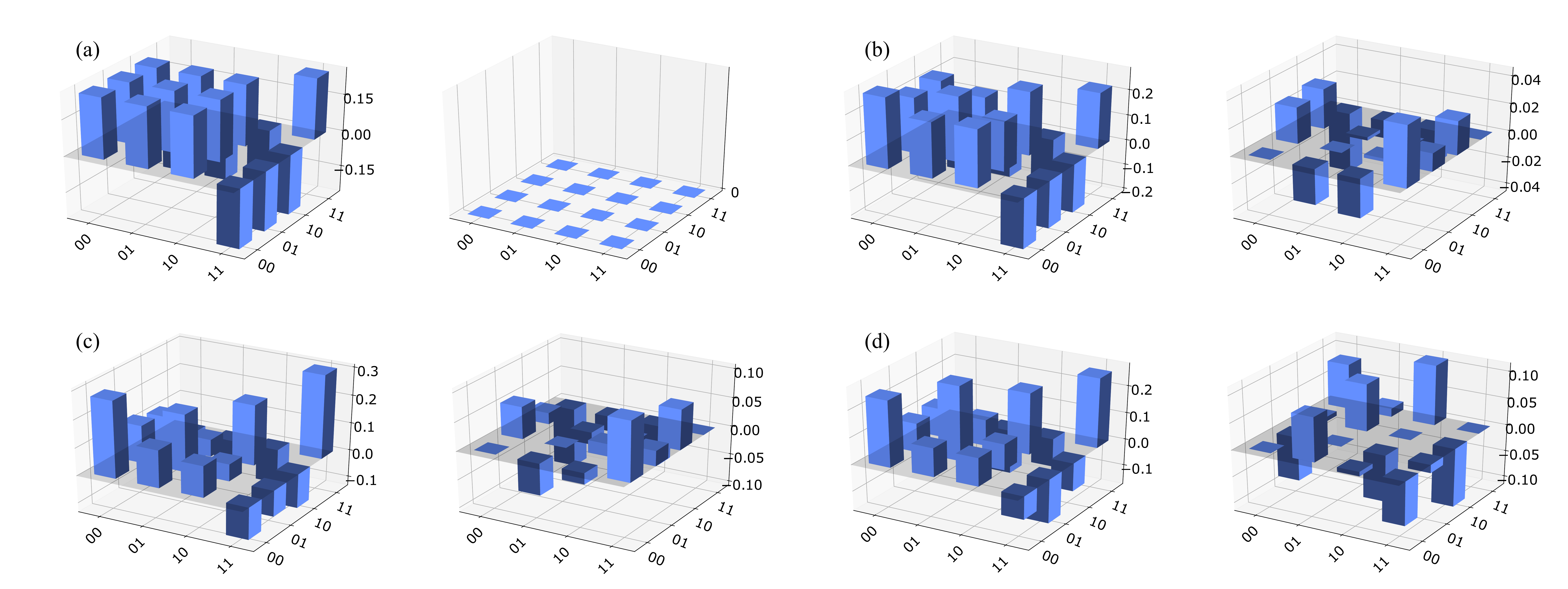}
		\caption{\label{fig:figure6_tomography} In each subfigure, the real part of the density matrix is on the left and the imaginary part is on the right. (a) Ideal density matrix of the 2-qubit cluster state. (b) Reconstructed density matrix of directly created 2-qubit linear cluster state using physical qubits $\{0,5\}$ on IBM Q 20 Tokyo. The density matrix is still close to the ideal case, in particular even though the imaginary density matrix elements are non-zero they all of the order $10^{-2}$. (c) and (d) are reconstructed density matrices of pairs 0 - 11 and 10 - 1 using MQNC.}
	\end{figure*}
        
    Before implementing the MBQC protocol to establish maximal entanglement between distant pair of qubits we investigate how the fidelity of the linear cluster state scales with its size. For IBM Q 20 Tokyo, we selected qubits $\{0,5,10,15,16\}$ while for IBM Q 20 Poughkeepsie, we selected qubits $\{5,0,1,2,3\}$ as seen in Fig.~\ref{fig:figure5_topology}(b). For state tomography, we performed five trials of 8192 shots each on $n$ qubits after the construction of an $n$-qubit linear cluster state $|G_n\rangle$. The obtained fidelities are shown in Table \ref{tab:nqubitLinearCluster}. For a sample result, a plot of the density matrix of the 2-qubit linear cluster state is shown in Fig. \ref{fig:figure6_tomography}(b).
        
    To establish direct maximal entanglement between distant qubits using the MBQC protocol, we picked qubits $\{0,5,6,11\}$ and $\{1,6,5,10\}$ for pairs $\{0,11\}$ and $\{1,10\}$, respectively. We performed single-qubit measurements in the $X$ basis on qubits $\{5,6\}$. Using one trial of post-selection for measurement results of `0' on both qubits and state-tomography, we find a state fidelity $F(\rho,|G_2\rangle)\approx0.70$ for pair $\{0,11\}$ and $F(\rho,|G_2\rangle)\approx0.63$ for pair $\{1,10\}$ compared with the ideal 2-qubit cluster state $|G_2\rangle = (|0+\rangle + |1-\rangle)/\sqrt{2}$.
    Note that $|G_2\rangle$ can be transformed into Bell pair $|\Phi^+\rangle$ via applying a Hadamard operation on either qubit.
    
    \begin{table}
        \centering
        \begin{tabular}{c|c|c}
            \hline
            $n$ qubits & IBM Q 20 Tokyo & IBM Q Poughkeepsie \\ \hline
            2 & 0.876 $\pm$ 0.005 & 0.821 $\pm$ 0.013 \\
            3 & 0.695 $\pm$ 0.003 & 0.700 $\pm$ 0.023 \\
            4 & 0.405 $\pm$ 0.003 & 0.595 $\pm$ 0.017 \\
            5 & 0.150 $\pm$ 0.010 & 0.373 $\pm$ 0.018\\ \hline
        \end{tabular}
        \caption{\label{tab:nqubitLinearCluster} Fidelity  $F(\rho,|G_n\rangle)$ of $n$-qubit linear cluster states on real devices.}
    \end{table}
        
    We see that the fidelities of the entangled pairs of qubits obtained via entanglement swapping and using MBQC on a linear cluster state are close to each other. This is not completely surprising. Preparing the initial resource for entanglement swapping requires two 2-qubit gates while to create a linear cluster state we reuire three 2-qubit gates. However, entanglement swapping proceeds by applying a Bell-state measurement which requires a further 2-qubit gate. On the other hand, the MBQC protocol requires only single-qubit measurements to proceed. 2-qubit gates are the main source of noise in both protocols and the total number of 2-qubit gates in both protocols is the same.

\section{\label{sec:QNConIBMQ}Quantum Network Coding on the IBM Q Experience}
    
    In the previous section, we investigated two protocols that can be used in the case when there is no contention for network resources. However, in real-world networks and in real physical systems, resource contention is inevitable and must be addressed. In this section we analyze an implementation of MQNC that is specifically designed to deal with resource contention in networks and in the systems such as IBM Q Experience devices.

    We implement the 2D 6-qubit cluster state part of the MQNC protocol (Step 2 in Fig.~\ref{fig:figure5_mqnc}) and evaluate the fidelity using state tomography on two remaining  2-qubit cluster states to the ideal quantum state,
    \begin{eqnarray} \label{eq:2crossingover}
        |G_{\times}\rangle & = &|G_2\rangle|G_2\rangle \nonumber\\
        & = & \frac{1}{2}(|0+\rangle + |1-\rangle) \otimes (|0+\rangle + |1-\rangle),
    \end{eqnarray}
    on qubit pairs $\{0,11\}$ and $\{1,10\}$. To confirm that the device implemented the protocol correctly transform it into two Bell pairs and perform three tests. We compute the correlation matrix, the concurrence and look for violation of the CHSH inequality for each qubit pair.

    \subsection{Implementation on IBM Q 20 Tokyo} \label{sec:implementation}
    
    Following Step 2 of the MQNC procedure in Fig.~\ref{fig:figure5_mqnc} to create $|G_{\times}\rangle$, the quantum circuit in Fig. \ref{fig:figure2_background}(c) was executed on IBM Q 20 Tokyo. For the 6-qubit cluster state, qubits $\{0,1,5,6,10,11\}$ were chosen in order to avoid the need for $\text{SWAP}$ gates. Under ideal conditions, qubits $\{0,11\}$ become maximally entangled and so do qubits $\{1,10\}$. The fidelity of this 4-qubit state was found to be $F(\rho,|G_{\times}\rangle)\approx0.41 \pm 0.01$ with twenty trials of state tomography where measurement result of qubits $\{5,6\}$ was post-selected to be `1'. The fidelity of the state on qubits $\{0,11\}$ was found to be $F(\rho,|G_2\rangle)\approx0.57 \pm 0.01$, and $F(\rho,|G_2\rangle)\approx0.58 \pm 0.01$ for qubits $\{1,10\}$.

    \subsection{Entanglement Verification} \label{entanglementVerify}

	\begin{figure}
		\centering
		\includegraphics[width=\columnwidth]{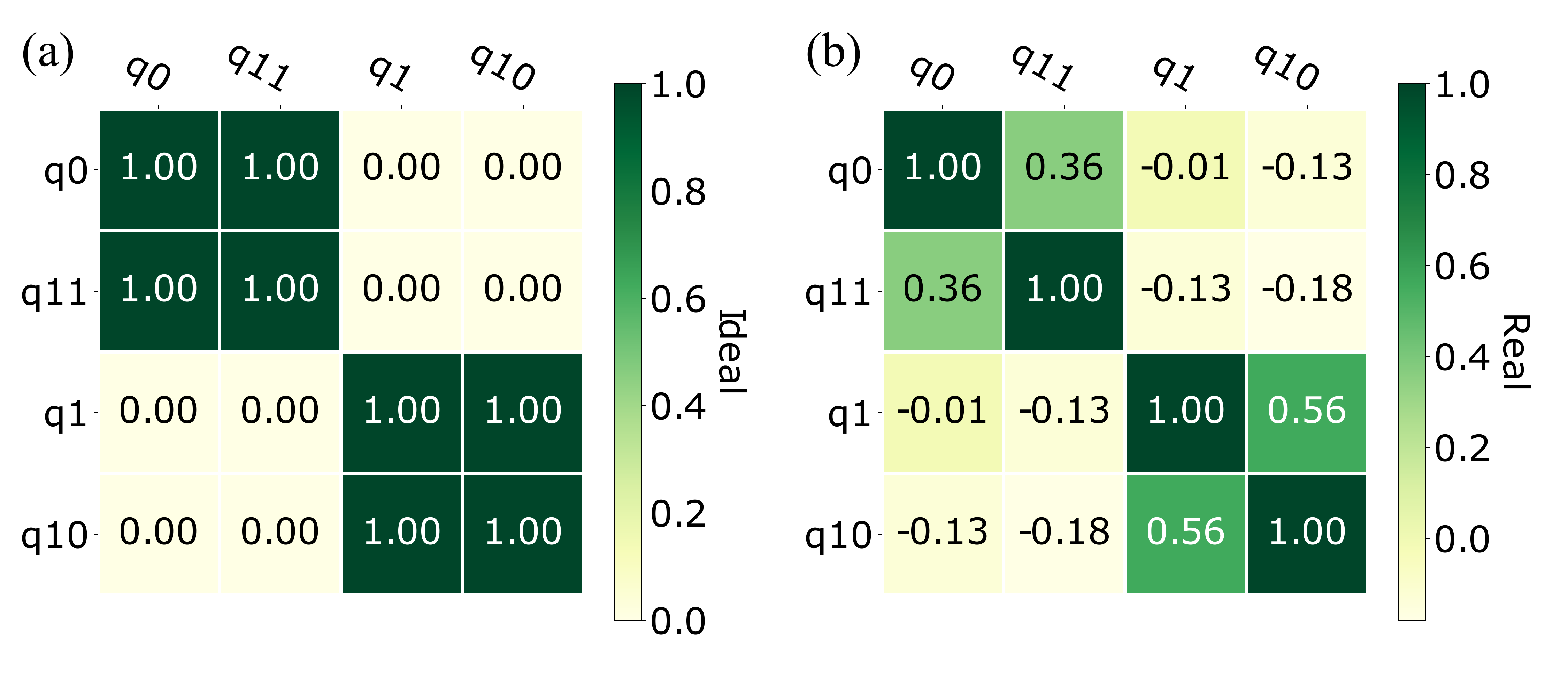}
		\caption{\label{fig:figure7_correlation} (a) Correlation matrix of the two 2-qubit cluster states using QASM simulator. (b) Correlation matrix for the same qubits obtained using the IBM Q 20 Tokyo. Standard deviation for the entangled and the non-entangled pairs was found to be 0.02.}
	\end{figure}
            
    To further analyze the procedure, we calculate the correlation matrix for the four remaining qubits by transforming each 2-qubit cluster state into Bell pair $|\Phi^+\rangle$. Fig. \ref{fig:figure7_correlation}(a), an ideal result using ten trials of the QASM simulator, shows that qubits $\{0,11\}$ are perfectly correlated and so are qubits $\{1,10\}$, and other entries are 0. This implies that there is no correlation between those qubits as one would expect.

    Fig.~\ref{fig:figure7_correlation}(b) is the correlation matrix using IBM Q 20 Tokyo with one hundred trials, showing a decline of correlation values from the ideal result. This should be expected from the fidelity of the state. However, the value for a pair expected to be entangled is significantly higher than a pair expected to be uncorrelated. The result from IBM Q 20 Tokyo reveals moderately large positive correlations where we expect them and significant negative correlations where we would hope for zero.
         
\begin{figure}
		\centering
		\includegraphics[width=\columnwidth]{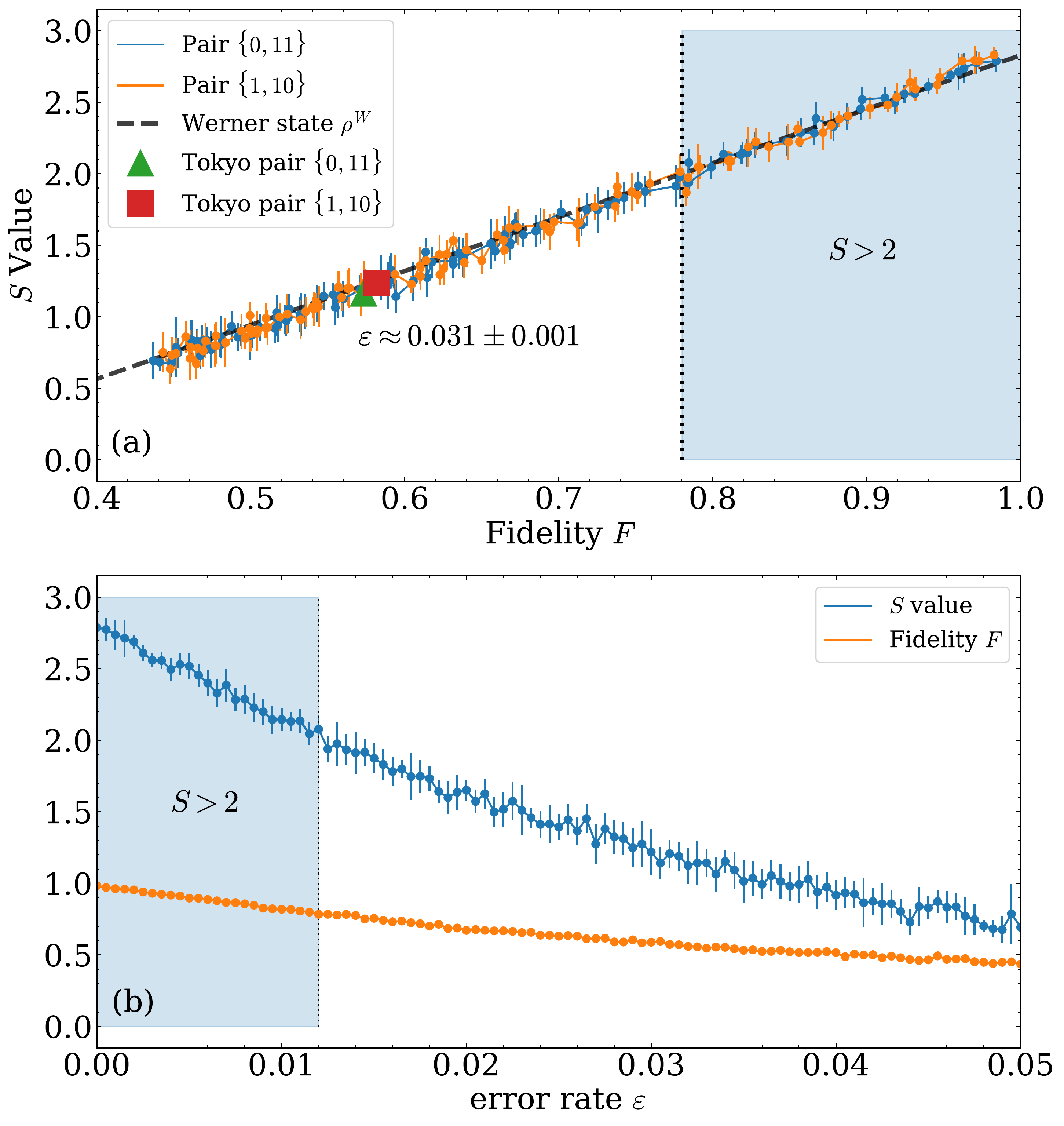}
		\caption{\label{fig:figure8_Svalue} (a) Plot of CHSH correlation value $S$ as a function of the fidelity $F$. The solid blue and orange lines were obtained using Qiskit's noisy simulator. They agree well with CHSH correlation values obtained using a Werner state $\rho^W$, represented by the dashed black line. The green triangle and red square represent data obtained using IBM Q 20 Tokyo for qubit pairs $\{0,11\}$ and $\{1,10\}$, respectively. The shaded region depicts parameter regime where the qubit pairs violate the CHSH inequality. (b) Plots of the CHSH correlation value $S$ and fidelity $F$ of qubit pair $\{0,11\}$ as functions of the error rate $\varepsilon$ using Qiskit's noisy simulator.}
	\end{figure}

    Seeing that the expected qubit pairs are correlated we now compute how entangled they are. A convenient measure to quantify the entanglement of 2-qubit mixed states is the concurrence $C(\rho)$ \cite{hill1997entanglement},
    \begin{equation}
        C(\rho) = \max\{0,\lambda_1-\lambda_2-\lambda_3-\lambda_4\},
    \end{equation}
    where $\lambda_1\dots\lambda_4$ are the eigenvalues of $R(\rho)$ in decreasing order, with
    \begin{equation}
        R(\rho)=\sqrt{\sqrt{\rho}\tilde{\rho}\sqrt{\rho}}, \qquad \tilde{\rho}=(Y\otimes Y)\rho^*(Y\otimes Y).
    \end{equation}
    We find that $C(\rho)=0.253\pm0.018$ for qubit pair $\{0,11\}$ (green triangle in Fig.~\ref{fig:figure8_Svalue}(a)) and $C(\rho)=0.312\pm0.021$ for qubit pair $\{1,10\}$ (red square in Fig.~\ref{fig:figure8_Svalue}(a)). Qubit pairs $\{0,1\}$, $\{0,10\}$, $\{1,11\}$ and $\{10,11\}$ are separable with vanishing concurrence as expected.
            
    Having established that the relevant qubit pairs do become entangled as the result of MQNC on IBM Q 20 Tokyo, a natural question to ask is how useful this entanglement actually is. One of the ultimate goals of quantum networks is to be able to produce nonlocal correlations between spatially-separated nodes that can be used to produce genuinely random cryptographic keys. To answer this question we perform the CHSH test \cite{CHSHOriginalPaper} on the entangled qubit pairs $\{0,11\}$ and $\{1,10\}$, respectively. The CHSH inequality is
    \begin{equation}\label{eq:CHSHeq}
        S=\langle A B\rangle-\left\langle A B^{\prime}\right\rangle+\left\langle A^{\prime} B\right\rangle+\left\langle A^{\prime} B^{\prime}\right\rangle \leq 2,
    \end{equation} 
    where $\langle O_1 O_2 \rangle=\text{Tr} \{O_1 O_2 \rho\}$, and $A=X$, $A'=Z$, $B=H$ and $B'=ZHZ$. Any state $\rho$ violating this inequality is said to be nonlocally correlated and can be used to produce certified random numbers that are guaranteed to be secure from adversaries limited by the no-signaling principle \cite{brunner2014bell}.
            
    Table \ref{tab:CHSH} shows the $S$ values from each pair using eight trials of 8192 shots each and post-selecting measurement outcomes `1' on qubits $\{1,4\}$. The $S$ value does not exceed 2, failing to demonstrate nonlocal correlations in the two 2-qubit cluster states. This result is to be expected from the states with fidelities obtained in Section \ref{sec:implementation} and is mainly the consequence of IBM Q 20 Tokyo being too noisy for protocols such as MQNC.
            
    \begin{table}
        \centering
        \begin{tabular}{c|c|c}
            \hline
            & pair & $S$ Value \\ \hline
            \multirow{2}{*}{Entangled pairs} & $\{0,11\}$ & 1.165 $\pm$ 0.04 \\ & $\{1,10\}$ & 1.235 $\pm$ 0.04 \\ \hline
            \multirow{4}{*}{Separable pairs} & $\{0,10\}$ & -0.405 $\pm$ 0.06 \\ & $\{0,1\}$ & -0.149 $\pm$ 0.03 \\
            & $\{10,11\}$ & -0.326 $\pm$ 0.04 \\ & $\{1,11\}$ & -0.084 $\pm$ 0.06 \\ \hline
        \end{tabular}
        \caption{\label{tab:CHSH} $S$ values of each pair using IBM Q 20 Tokyo with 8 trials. Pairs $\{0,11\}$ and $\{1,10\}$ show higher $S$ values compared to the separable pairs. However these $S$ values are not high enough to demonstrate nonlocal correlations.}
    \end{table}

    We now consider a simple noise model that allows us to extract the required error rates which would allow MQNC to produce qubit pairs that violate the CHSH inequality. We model the final mixed state of the two entangled qubits as a rotated Werner state \cite{WernerOriginalPaper},
    \begin{equation} \label{eq:werner}
        \rho^W = \frac{4F - 1}{3}\left|G_2\right\rangle\left\langle G_2\right| + \frac{1-F}{3}I,
    \end{equation}
    where $I$ is the identity operator and $F=F(\rho^W,|G_2\rangle)$ is the fidelity with respect to $|G_2\rangle$. The rotated Werner state $\rho^W$ violates the CHSH inequality when $F \gtrsim 0.78$ \cite{WernerFidelityCHSH}, as can be seen in Fig.~\ref{fig:figure8_Svalue}(a).
            
    By using Qiskit's noisy simulator, we vary the probability of an error $\varepsilon$ of single-qubit operation $O$ using the depolarizing channel,
    \begin{equation}
        E_O(\rho) = (1 - \varepsilon)O\rho O^{\dagger} + \varepsilon\frac{I}{2},
    \end{equation}
    from 0 to 0.05 (101 data points repeated 10 times each and 1024 shots per time). We model the total error on 2-qubit gates as two independent errors on the control and target qubits. Despite the simplicity of this error model we see in Fig.~\ref{fig:figure8_Svalue}(a) that it agrees with the experimental data obtained from IBM Q 20 Tokyo. Both Qiskit's noisy simulation (blue and orange lines) and experimental data (green triangle and red square) obtained from IBM Q 20 Tokyo agree well with prediction obtained using a Werner state $\rho^W$ (black dashed line). To make this agreement more quantitative, we use the fidelities $F(\rho,|G_2\rangle)$ of pairs $\{0,11\}$ and $\{1,10\}$ to compute the respective rotated Werner states. We use $\rho^W_1$ to denote the rotated Werner state corresponding to pair $\{0,11\}$ and $\rho^W_2$ to denote the rotated Werner state corresponding to pair $\{1,10\}$. The fidelity of the two 2-qubit cluster states with respect to these rotated Werner states were found to be $F(\rho,\rho^W_1)=0.955\pm0.003$ for pair $\{0,11\}$ and $F(\rho,\rho^W_2)=0.943\pm0.005$ for pair $\{1,10\}$. An advantage of this simple noise model is that it allows us to treat the CHSH correlation value $S$ and the fidelity of the final qubit pairs as being functions of single-qubit error rate $\varepsilon$ only as shown in Fig.~\ref{fig:figure8_Svalue}(b) which allows us to extract the error rates required for CHSH violation as shown below.
            
    We simulate the 6-qubit cluster state part of MQNC twice, once to perform state tomography for 2 entangled qubits and another time for the CHSH experiment on the same entangled qubit pairs. The result from the simulation is sorted by fidelity and plotted against the $S$ value from CHSH experiment in Fig.~\ref{fig:figure8_Svalue}(a). It shows that the noise model in Eq.~(\ref{eq:werner}) is adequate because both data points from IBM Q 20 Tokyo are on the predicted curve using the noisy QASM simulator. The two data points do not have sufficiently high fidelity to violate the CHSH inequality. Our simple noise model readily allows us to extrapolate the critical single-qubit error rate $\varepsilon_{crit}$ which results in a CHSH violation. Fig.~\ref{fig:figure8_Svalue}(b) suggest that $\varepsilon_{crit}\approx1.2\%$ which is less than half of the depolarising error rate of the 2 data points obtained from IBM Q 20 Tokyo.
        
    Experimentally, we have found that MQNC is not yet achievable on IBM Q 20 Tokyo.  After the completion of MQNC, we should be left with two independent 2-qubit cluster states, but the six-qubit entanglement may or may not give us complete independence due to imperfections in the state creation and measurement. To check, in addition to calculating the CHSH correlation value $S$ for the qubit pairs we \emph{expect} to be entangled, we also calculated $S$ for qubit pairs we expect to \emph{not} be correlated--each term in Eq.~(\ref{eq:CHSHeq}) should vanish if the two qubits are completely independent. The values in Fig.~\ref{fig:figure7_correlation}(b) and Table~\ref{tab:CHSH} suggest that residual correlations remain after post-selection. Of course, post-selection only emulates the full behavior of MQNC; actual measurement and feed-forward would produce different results, but our data here is suggestive.
    
    On the other hand we can see that the fidelities for the final entangled pairs are $F(\rho,|G_2\rangle)>0.5$. This means that the entangled pairs can be in principle further purified to yield entangled pairs with higher fidelities, limited only by the fidelity of local gates and measurements at repeater nodes.
        
    Poughkeepsie, a newer machine, is superior to Tokyo in terms of fidelity but its interconnection topology is not rich enough to directly implement MQNC. The rapid improvements in hardware suggest that MQNC will violate the CHSH inequality within a generation or two, if the processor topology allows.

\section{\label{sec:Discussion}Discussion}
    
    In this paper, we have taken a step from the theory of quantum network coding toward practical use on real devices. By implementing and comparing to more traditional entanglement swapping and linear cluster states, we can assess the conditions under which each of the three approaches will best serve applications. This will benefit multiple applications competing for access to a network, as well as help us to coordinate use of resources inside a single quantum system as part of the algorithm compilation and optimization process. Of course, a common usage scenario in networks is to assume each link is a long-distance optical channel and we are implementing QNC in a distributed fashion. An alternative use inside a single system might be switching longer-distance connections when the system is used as a quantum router in a network.
        
    This paper has made several contributions: it has demonstrated how a quantum computer can be used to model an entangling quantum network; experimentally confirmed the operation of quantum network coding in a superconducting device; shown that a Werner state error model is sufficient for modeling errors in cluster states even in the presence of biased noise; and examined the possibility of replacing qubit swapping with linear cluster states or QNC to move qubits around within systems.

    Entangling quantum networks are under development.  An important tool in development is simulation.  That simulation can be executed using a classical computer, as has been commonly done in quantum repeater design.  Here, we have shown that this simulation can be performed using a quantum computer, for certain parameter values.  The next steps toward improving the usefulness of this approach are to simulate stochastic behavior of photon loss with and without competing traffic on a large network.

    Quantum network coding, building on classical network coding, is widely accepted as a functional protocol.  However, to establish its utility in real-world operation, we must implement and analyze its behavior with respect to the imperfections of real systems.  In this paper, we have shown that, with appropriately compiled algorithms, a one-qubit error rate of about 1.2\% will allow the protocol to reach an $S$ value of 2. This would in turn allow for extraction of certified random keys are a finite rate. The current hardware error rate across the set of qubits used in our protocol corresponds roughly to $\varepsilon = 3.1\pm 0.1\%$.  This error model supplied with Qiskit will model Werner states, and we can see that the achieved fidelity and $S$ value correspond well with the Werner model simulations.

    Finally, an important factor in compiling algorithms for quantum computers is the assignment of program variables to locations in the system, and their movement within the system via SWAP gates as necessary to couple with other qubits.  We propose as an alternative that a cluster state can be used to move qubits, incorporating single-qubit rotations in the process.  Of course, this approach assumes that the needed interstitial qubits are otherwise idle. Further evaluation of the utility of both linear graph-based teleportation and QNC on a monolithic quantum computer await the arrival of feed-forward functionality. This points the way to hybrid measurement/gate-based quantum computation.

\section*{Acknowledgement}

    This material is based upon work supported by the Air Force Office of Scientific Research under award number FA2386-19-1-4038. The results presented in this paper were obtained in part using an IBM Q quantum computing system as part of the IBM Q Network. The views expressed are those of the authors and do not reflect the official policy or position of IBM or the IBM Q team.
        
    During the submission of our work we became aware of the following optical implementation of quantum network coding \cite{lu2019experimental} based on a different protocol building on preshared entanglement \cite{hayashi2007prior}.

\appendix
\section{Qiskit Version}

    The version of Qiskit packages we use are listed in Table \ref{tab:QiskitVer}.
    \begin{table}[h!]
        \centering
        \begin{tabular}{c|c}
        \hline
            name & version \\ \hline
            qiskit & 0.10.5 \\
            qiskit-terra & 0.8.2 \\
            qiskit-ignis & 0.1.1 \\
            qiskit-aer & 0.2.1 \\
            qiskit-ibmq-provider & 0.2.2 \\
            qiskit-aqua & 0.5.2 \\ \hline
        \end{tabular}
        \caption{\label{tab:QiskitVer} Qiskit packages version}
    \end{table}

\section{Date-time}

    Each experiment was performed on the dates listed in Table \ref{tab:ExperimentDate}. 
    \begin{table}[h!]
        \centering
        \begin{tabular}{m{0.3\textwidth}|c}
        \hline
            Experiment & Date-time \\ \hline
            4 qubit Entanglement Swapping and 4 qubit Linear Cluster state on IBM Q 20 Tokyo & 2019/7/24 \\ \hline
            $n$-qubit Linear Cluster on IBM Q 20 Tokyo & 2019/8/21-22\\ \hline 
            $n$-qubit Linear Cluster on IBM Q Poughkeepsie & 2019/8/27-28 \\ \hline
            6 qubit cluster state fidelity on IBM Q 20 Tokyo & 2019/8/25-27 \\ \hline
            Correlation Matrix of 2 crossing-over pairs on IBM Q 20 Tokyo & 2019/8/27 \\ \hline
            CHSH inequality of 2 crossing-over pairs on IBM Q 20 Tokyo & 2019/8/24 \\ \hline
            
        \end{tabular}
        \caption{Date and time when experimental data have been taken}
        \label{tab:ExperimentDate}
    \end{table}

%\bibliography{reference}

\begin{thebibliography}{69}%
	\makeatletter
	\providecommand \@ifxundefined [1]{%
		\@ifx{#1\undefined}
	}%
	\providecommand \@ifnum [1]{%
		\ifnum #1\expandafter \@firstoftwo
		\else \expandafter \@secondoftwo
		\fi
	}%
	\providecommand \@ifx [1]{%
		\ifx #1\expandafter \@firstoftwo
		\else \expandafter \@secondoftwo
		\fi
	}%
	\providecommand \natexlab [1]{#1}%
	\providecommand \enquote  [1]{``#1''}%
	\providecommand \bibnamefont  [1]{#1}%
	\providecommand \bibfnamefont [1]{#1}%
	\providecommand \citenamefont [1]{#1}%
	\providecommand \href@noop [0]{\@secondoftwo}%
	\providecommand \href [0]{\begingroup \@sanitize@url \@href}%
	\providecommand \@href[1]{\@@startlink{#1}\@@href}%
	\providecommand \@@href[1]{\endgroup#1\@@endlink}%
	\providecommand \@sanitize@url [0]{\catcode `\\12\catcode `\$12\catcode
		`\&12\catcode `\#12\catcode `\^12\catcode `\_12\catcode `\%12\relax}%
	\providecommand \@@startlink[1]{}%
	\providecommand \@@endlink[0]{}%
	\providecommand \url  [0]{\begingroup\@sanitize@url \@url }%
	\providecommand \@url [1]{\endgroup\@href {#1}{\urlprefix }}%
	\providecommand \urlprefix  [0]{URL }%
	\providecommand \Eprint [0]{\href }%
	\providecommand \doibase [0]{http://dx.doi.org/}%
	\providecommand \selectlanguage [0]{\@gobble}%
	\providecommand \bibinfo  [0]{\@secondoftwo}%
	\providecommand \bibfield  [0]{\@secondoftwo}%
	\providecommand \translation [1]{[#1]}%
	\providecommand \BibitemOpen [0]{}%
	\providecommand \bibitemStop [0]{}%
	\providecommand \bibitemNoStop [0]{.\EOS\space}%
	\providecommand \EOS [0]{\spacefactor3000\relax}%
	\providecommand \BibitemShut  [1]{\csname bibitem#1\endcsname}%
	\let\auto@bib@innerbib\@empty
	%</preamble>
	\bibitem [{\citenamefont {Wehner}\ \emph {et~al.}(2018)\citenamefont {Wehner},
		\citenamefont {Elkouss},\ and\ \citenamefont {Hanson}}]{Wehner18:eaam9288}%
	\BibitemOpen
	\bibfield  {author} {\bibinfo {author} {\bibfnamefont {S.}~\bibnamefont
			{Wehner}}, \bibinfo {author} {\bibfnamefont {D.}~\bibnamefont {Elkouss}}, \
		and\ \bibinfo {author} {\bibfnamefont {R.}~\bibnamefont {Hanson}},\ }\href
	{\doibase 10.1126/science.aam9288} {\bibfield  {journal} {\bibinfo  {journal}
			{Science}\ }\textbf {\bibinfo {volume} {362}} (\bibinfo {year} {2018}),\
		10.1126/science.aam9288},\ \Eprint
	{http://arxiv.org/abs/\href{http://science.sciencemag.org/content/362/6412/eaam9288.full.pdf}}
	{\href{http://science.sciencemag.org/content/362/6412/eaam9288.full.pdf}}
	\BibitemShut {NoStop}%
	\bibitem [{\citenamefont {Kimble}(2008)}]{kimble08:_quant_internet}%
	\BibitemOpen
	\bibfield  {author} {\bibinfo {author} {\bibfnamefont {H.~J.}\ \bibnamefont
			{Kimble}},\ }\href@noop {} {\bibfield  {journal} {\bibinfo  {journal}
			{Nature}\ }\textbf {\bibinfo {volume} {453}},\ \bibinfo {pages} {1023}
		(\bibinfo {year} {2008})}\BibitemShut {NoStop}%
	\bibitem [{\citenamefont {Muralidharan}\ \emph {et~al.}(2016)\citenamefont
		{Muralidharan}, \citenamefont {Li}, \citenamefont {Kim}, \citenamefont
		{L{\"u}tkenhaus}, \citenamefont {Lukin},\ and\ \citenamefont
		{Jiang}}]{muralidharan2016optimal}%
	\BibitemOpen
	\bibfield  {author} {\bibinfo {author} {\bibfnamefont {S.}~\bibnamefont
			{Muralidharan}}, \bibinfo {author} {\bibfnamefont {L.}~\bibnamefont {Li}},
		\bibinfo {author} {\bibfnamefont {J.}~\bibnamefont {Kim}}, \bibinfo {author}
		{\bibfnamefont {N.}~\bibnamefont {L{\"u}tkenhaus}}, \bibinfo {author}
		{\bibfnamefont {M.~D.}\ \bibnamefont {Lukin}}, \ and\ \bibinfo {author}
		{\bibfnamefont {L.}~\bibnamefont {Jiang}},\ }\href
	{https://www.nature.com/articles/srep20463} {\bibfield  {journal} {\bibinfo
			{journal} {Scientific Reports}\ }\textbf {\bibinfo {volume} {6}},\ \bibinfo
		{pages} {20463} (\bibinfo {year} {2016})}\BibitemShut {NoStop}%
	\bibitem [{\citenamefont {Van{
			}Meter}(2014)}]{van-meter14:_quantum_networking}%
	\BibitemOpen
	\bibfield  {author} {\bibinfo {author} {\bibfnamefont {R.}~\bibnamefont {Van{
				}Meter}},\ }\href@noop {} {\emph {\bibinfo {title} {Quantum Networking}}}\
	(\bibinfo  {publisher} {Wiley-ISTE},\ \bibinfo {year} {2014})\BibitemShut
	{NoStop}%
	\bibitem [{\citenamefont {Gisin}\ and\ \citenamefont
		{Thew}(2007)}]{gisin2007quantum}%
	\BibitemOpen
	\bibfield  {author} {\bibinfo {author} {\bibfnamefont {N.}~\bibnamefont
			{Gisin}}\ and\ \bibinfo {author} {\bibfnamefont {R.}~\bibnamefont {Thew}},\
	}\href@noop {} {\bibfield  {journal} {\bibinfo  {journal} {Nature Photonics}\
		}\textbf {\bibinfo {volume} {1}},\ \bibinfo {pages} {165} (\bibinfo {year}
		{2007})}\BibitemShut {NoStop}%
	\bibitem [{\citenamefont {Bennett}\ and\ \citenamefont
		{Brassard}(1984)}]{bennett:bb84}%
	\BibitemOpen
	\bibfield  {author} {\bibinfo {author} {\bibfnamefont {C.~H.}\ \bibnamefont
			{Bennett}}\ and\ \bibinfo {author} {\bibfnamefont {G.}~\bibnamefont
			{Brassard}},\ }in\ \href
	{http://www.research.ibm.com/people/b/bennetc/bennettc198469790513.pdf}
	{\emph {\bibinfo {booktitle} {Proc. IEEE International Conference on
				Computers, Systems, and Signal Processing}}}\ (\bibinfo  {publisher} {IEEE},\
	\bibinfo {year} {1984})\ pp.\ \bibinfo {pages} {175--179}\BibitemShut
	{NoStop}%
	\bibitem [{\citenamefont {Ekert}(1991)}]{ekert1991qcb}%
	\BibitemOpen
	\bibfield  {author} {\bibinfo {author} {\bibfnamefont {A.}~\bibnamefont
			{Ekert}},\ }\href@noop {} {\bibfield  {journal} {\bibinfo  {journal}
			{Physical Review Letters}\ }\textbf {\bibinfo {volume} {67}},\ \bibinfo
		{pages} {661} (\bibinfo {year} {1991})}\BibitemShut {NoStop}%
	\bibitem [{\citenamefont {Elliott}\ \emph {et~al.}(2003)\citenamefont
		{Elliott}, \citenamefont {Pearson},\ and\ \citenamefont
		{Troxel}}]{elliott:qkd-net}%
	\BibitemOpen
	\bibfield  {author} {\bibinfo {author} {\bibfnamefont {C.}~\bibnamefont
			{Elliott}}, \bibinfo {author} {\bibfnamefont {D.}~\bibnamefont {Pearson}}, \
		and\ \bibinfo {author} {\bibfnamefont {G.}~\bibnamefont {Troxel}},\ }in\
	\href@noop {} {\emph {\bibinfo {booktitle} {Proc. SIGCOMM 2003}}},\ \bibinfo
	{organization} {ACM}\ (\bibinfo  {publisher} {ACM},\ \bibinfo {year}
	{2003})\BibitemShut {NoStop}%
	\bibitem [{\citenamefont {Gisin}\ \emph {et~al.}(2002)\citenamefont {Gisin},
		\citenamefont {Ribordy}, \citenamefont {Tittel},\ and\ \citenamefont
		{Zbinden}}]{gisin2002quantum}%
	\BibitemOpen
	\bibfield  {author} {\bibinfo {author} {\bibfnamefont {N.}~\bibnamefont
			{Gisin}}, \bibinfo {author} {\bibfnamefont {G.}~\bibnamefont {Ribordy}},
		\bibinfo {author} {\bibfnamefont {W.}~\bibnamefont {Tittel}}, \ and\ \bibinfo
		{author} {\bibfnamefont {H.}~\bibnamefont {Zbinden}},\ }\href@noop {}
	{\bibfield  {journal} {\bibinfo  {journal} {Reviews of Modern Physics}\
		}\textbf {\bibinfo {volume} {74}},\ \bibinfo {pages} {145} (\bibinfo {year}
		{2002})}\BibitemShut {NoStop}%
	\bibitem [{\citenamefont {Buhrman}\ and\ \citenamefont
		{R\"ohrig}(2003)}]{buhrman03:_dist_qc}%
	\BibitemOpen
	\bibfield  {author} {\bibinfo {author} {\bibfnamefont {H.}~\bibnamefont
			{Buhrman}}\ and\ \bibinfo {author} {\bibfnamefont {H.}~\bibnamefont
			{R\"ohrig}},\ }\enquote {\bibinfo {title} {Mathematical foundations of
			computer science 2003},}\ \ (\bibinfo  {publisher} {Springer-Verlag},\
	\bibinfo {year} {2003})\ Chap.\ \bibinfo {chapter} {Distributed Quantum
		Computing}, pp.\ \bibinfo {pages} {1--20}\BibitemShut {NoStop}%
	\bibitem [{\citenamefont {Van~Meter}\ and\ \citenamefont
		{Devitt}(2016)}]{van2016path}%
	\BibitemOpen
	\bibfield  {author} {\bibinfo {author} {\bibfnamefont {R.}~\bibnamefont
			{Van~Meter}}\ and\ \bibinfo {author} {\bibfnamefont {S.~J.}\ \bibnamefont
			{Devitt}},\ }\href@noop {} {\bibfield  {journal} {\bibinfo  {journal}
			{Computer}\ }\textbf {\bibinfo {volume} {49}},\ \bibinfo {pages} {31}
		(\bibinfo {year} {2016})}\BibitemShut {NoStop}%
	\bibitem [{\citenamefont {Broadbent}\ \emph {et~al.}(2009)\citenamefont
		{Broadbent}, \citenamefont {Fitzsimons},\ and\ \citenamefont
		{Kashefi}}]{broadbent2009universal}%
	\BibitemOpen
	\bibfield  {author} {\bibinfo {author} {\bibfnamefont {A.}~\bibnamefont
			{Broadbent}}, \bibinfo {author} {\bibfnamefont {J.}~\bibnamefont
			{Fitzsimons}}, \ and\ \bibinfo {author} {\bibfnamefont {E.}~\bibnamefont
			{Kashefi}},\ }in\ \href@noop {} {\emph {\bibinfo {booktitle} {2009 50th
				Annual IEEE Symposium on Foundations of Computer Science}}}\ (\bibinfo
	{organization} {IEEE},\ \bibinfo {year} {2009})\ pp.\ \bibinfo {pages}
	{517--526}\BibitemShut {NoStop}%
	\bibitem [{\citenamefont {Hajdu{\v{s}}ek}\ \emph {et~al.}(2015)\citenamefont
		{Hajdu{\v{s}}ek}, \citenamefont {P{\'e}rez-Delgado},\ and\ \citenamefont
		{Fitzsimons}}]{hajduvsek2015device}%
	\BibitemOpen
	\bibfield  {author} {\bibinfo {author} {\bibfnamefont {M.}~\bibnamefont
			{Hajdu{\v{s}}ek}}, \bibinfo {author} {\bibfnamefont {C.~A.}\ \bibnamefont
			{P{\'e}rez-Delgado}}, \ and\ \bibinfo {author} {\bibfnamefont {J.~F.}\
			\bibnamefont {Fitzsimons}},\ }\href@noop {} {\bibfield  {journal} {\bibinfo
			{journal} {arXiv:1502.02563}\ } (\bibinfo {year} {2015})}\BibitemShut
	{NoStop}%
	\bibitem [{\citenamefont {Morimae}\ and\ \citenamefont
		{Fujii}(2013)}]{morimae2013blind}%
	\BibitemOpen
	\bibfield  {author} {\bibinfo {author} {\bibfnamefont {T.}~\bibnamefont
			{Morimae}}\ and\ \bibinfo {author} {\bibfnamefont {K.}~\bibnamefont
			{Fujii}},\ }\href@noop {} {\bibfield  {journal} {\bibinfo  {journal}
			{Physical Review A}\ }\textbf {\bibinfo {volume} {87}},\ \bibinfo {pages}
		{050301} (\bibinfo {year} {2013})}\BibitemShut {NoStop}%
	\bibitem [{\citenamefont {Hayashi}\ and\ \citenamefont
		{Hajdu{\v{s}}ek}(2018)}]{hayashi2018self}%
	\BibitemOpen
	\bibfield  {author} {\bibinfo {author} {\bibfnamefont {M.}~\bibnamefont
			{Hayashi}}\ and\ \bibinfo {author} {\bibfnamefont {M.}~\bibnamefont
			{Hajdu{\v{s}}ek}},\ }\href@noop {} {\bibfield  {journal} {\bibinfo  {journal}
			{Physical Review A}\ }\textbf {\bibinfo {volume} {97}},\ \bibinfo {pages}
		{052308} (\bibinfo {year} {2018})}\BibitemShut {NoStop}%
	\bibitem [{\citenamefont {Briegel}\ \emph {et~al.}(1998)\citenamefont
		{Briegel}, \citenamefont {D{\"u}r}, \citenamefont {Cirac},\ and\
		\citenamefont {Zoller}}]{briegel1998quantum}%
	\BibitemOpen
	\bibfield  {author} {\bibinfo {author} {\bibfnamefont {H.-J.}\ \bibnamefont
			{Briegel}}, \bibinfo {author} {\bibfnamefont {W.}~\bibnamefont {D{\"u}r}},
		\bibinfo {author} {\bibfnamefont {J.~I.}\ \bibnamefont {Cirac}}, \ and\
		\bibinfo {author} {\bibfnamefont {P.}~\bibnamefont {Zoller}},\ }\href@noop {}
	{\bibfield  {journal} {\bibinfo  {journal} {Physical Review Letters}\
		}\textbf {\bibinfo {volume} {81}},\ \bibinfo {pages} {5932} (\bibinfo {year}
		{1998})}\BibitemShut {NoStop}%
	\bibitem [{\citenamefont {D\"ur}\ \emph {et~al.}(1999)\citenamefont {D\"ur},
		\citenamefont {Briegel}, \citenamefont {Cirac},\ and\ \citenamefont
		{Zoller}}]{DurQuantumRepeater}%
	\BibitemOpen
	\bibfield  {author} {\bibinfo {author} {\bibfnamefont {W.}~\bibnamefont
			{D\"ur}}, \bibinfo {author} {\bibfnamefont {H.-J.}\ \bibnamefont {Briegel}},
		\bibinfo {author} {\bibfnamefont {J.~I.}\ \bibnamefont {Cirac}}, \ and\
		\bibinfo {author} {\bibfnamefont {P.}~\bibnamefont {Zoller}},\ }\href
	{\doibase 10.1103/PhysRevA.59.169} {\bibfield  {journal} {\bibinfo  {journal}
			{Phys. Rev. A}\ }\textbf {\bibinfo {volume} {59}},\ \bibinfo {pages} {169}
		(\bibinfo {year} {1999})}\BibitemShut {NoStop}%
	\bibitem [{\citenamefont {Zwerger}\ \emph {et~al.}(2018)\citenamefont
		{Zwerger}, \citenamefont {Pirker}, \citenamefont {Dunjko}, \citenamefont
		{Briegel},\ and\ \citenamefont {D\"ur}}]{PirkerLong-Range}%
	\BibitemOpen
	\bibfield  {author} {\bibinfo {author} {\bibfnamefont {M.}~\bibnamefont
			{Zwerger}}, \bibinfo {author} {\bibfnamefont {A.}~\bibnamefont {Pirker}},
		\bibinfo {author} {\bibfnamefont {V.}~\bibnamefont {Dunjko}}, \bibinfo
		{author} {\bibfnamefont {H.~J.}\ \bibnamefont {Briegel}}, \ and\ \bibinfo
		{author} {\bibfnamefont {W.}~\bibnamefont {D\"ur}},\ }\href {\doibase
		10.1103/PhysRevLett.120.030503} {\bibfield  {journal} {\bibinfo  {journal}
			{Phys. Rev. Lett.}\ }\textbf {\bibinfo {volume} {120}},\ \bibinfo {pages}
		{030503} (\bibinfo {year} {2018})}\BibitemShut {NoStop}%
	\bibitem [{\citenamefont {Jiang}\ \emph {et~al.}(2009)\citenamefont {Jiang},
		\citenamefont {Taylor}, \citenamefont {Nemoto}, \citenamefont {Munro},
		\citenamefont {Van{ }Meter},\ and\ \citenamefont
		{Lukin}}]{PhysRevA.79.032325}%
	\BibitemOpen
	\bibfield  {author} {\bibinfo {author} {\bibfnamefont {L.}~\bibnamefont
			{Jiang}}, \bibinfo {author} {\bibfnamefont {J.~M.}\ \bibnamefont {Taylor}},
		\bibinfo {author} {\bibfnamefont {K.}~\bibnamefont {Nemoto}}, \bibinfo
		{author} {\bibfnamefont {W.~J.}\ \bibnamefont {Munro}}, \bibinfo {author}
		{\bibfnamefont {R.}~\bibnamefont {Van{ }Meter}}, \ and\ \bibinfo {author}
		{\bibfnamefont {M.~D.}\ \bibnamefont {Lukin}},\ }\href {\doibase
		10.1103/PhysRevA.79.032325} {\bibfield  {journal} {\bibinfo  {journal} {Phys.
				Rev. A}\ }\textbf {\bibinfo {volume} {79}},\ \bibinfo {pages} {032325}
		(\bibinfo {year} {2009})}\BibitemShut {NoStop}%
	\bibitem [{\citenamefont {Fowler}\ \emph {et~al.}(2010)\citenamefont {Fowler},
		\citenamefont {Wang}, \citenamefont {Hill}, \citenamefont {Ladd},
		\citenamefont {Van{ }Meter},\ and\ \citenamefont
		{Hollenberg}}]{PhysRevLett.104.180503}%
	\BibitemOpen
	\bibfield  {author} {\bibinfo {author} {\bibfnamefont {A.~G.}\ \bibnamefont
			{Fowler}}, \bibinfo {author} {\bibfnamefont {D.~S.}\ \bibnamefont {Wang}},
		\bibinfo {author} {\bibfnamefont {C.~D.}\ \bibnamefont {Hill}}, \bibinfo
		{author} {\bibfnamefont {T.~D.}\ \bibnamefont {Ladd}}, \bibinfo {author}
		{\bibfnamefont {R.}~\bibnamefont {Van{ }Meter}}, \ and\ \bibinfo {author}
		{\bibfnamefont {L.~C.~L.}\ \bibnamefont {Hollenberg}},\ }\href {\doibase
		10.1103/PhysRevLett.104.180503} {\bibfield  {journal} {\bibinfo  {journal}
			{Phys. Rev. Lett.}\ }\textbf {\bibinfo {volume} {104}},\ \bibinfo {pages}
		{180503} (\bibinfo {year} {2010})}\BibitemShut {NoStop}%
	\bibitem [{\citenamefont {Devitt}\ \emph {et~al.}(2013)\citenamefont {Devitt},
		\citenamefont {Munro},\ and\ \citenamefont {Nemoto}}]{devitt13:rpp-qec}%
	\BibitemOpen
	\bibfield  {author} {\bibinfo {author} {\bibfnamefont {S.~J.}\ \bibnamefont
			{Devitt}}, \bibinfo {author} {\bibfnamefont {W.~J.}\ \bibnamefont {Munro}}, \
		and\ \bibinfo {author} {\bibfnamefont {K.}~\bibnamefont {Nemoto}},\ }\href
	{http://stacks.iop.org/0034-4885/76/i=7/a=076001} {\bibfield  {journal}
		{\bibinfo  {journal} {Reports on Progress in Physics}\ }\textbf {\bibinfo
			{volume} {76}},\ \bibinfo {pages} {076001} (\bibinfo {year}
		{2013})}\BibitemShut {NoStop}%
	\bibitem [{\citenamefont {D\"ur}\ \emph {et~al.}(2003)\citenamefont {D\"ur},
		\citenamefont {Aschauer},\ and\ \citenamefont {Briegel}}]{dur2003mep}%
	\BibitemOpen
	\bibfield  {author} {\bibinfo {author} {\bibfnamefont {W.}~\bibnamefont
			{D\"ur}}, \bibinfo {author} {\bibfnamefont {H.}~\bibnamefont {Aschauer}}, \
		and\ \bibinfo {author} {\bibfnamefont {H.-J.}\ \bibnamefont {Briegel}},\
	}\href {\doibase 10.1103/PhysRevLett.91.107903} {\bibfield  {journal}
		{\bibinfo  {journal} {Phys. Rev. Lett.}\ }\textbf {\bibinfo {volume} {91}},\
		\bibinfo {pages} {107903} (\bibinfo {year} {2003})}\BibitemShut {NoStop}%
	\bibitem [{\citenamefont {Dahlberg}\ \emph {et~al.}(2019)\citenamefont
		{Dahlberg}, \citenamefont {Skrzypczyk}, \citenamefont {Coopmans},
		\citenamefont {Wubben}, \citenamefont {Rozp{\k{e}}dek}, \citenamefont
		{Pompili}, \citenamefont {Stolk}, \citenamefont {Pawe{\l}czak}, \citenamefont
		{Knegjens}, \citenamefont {Hanson} \emph {et~al.}}]{dahlberg2019link}%
	\BibitemOpen
	\bibfield  {author} {\bibinfo {author} {\bibfnamefont {A.}~\bibnamefont
			{Dahlberg}}, \bibinfo {author} {\bibfnamefont {M.}~\bibnamefont
			{Skrzypczyk}}, \bibinfo {author} {\bibfnamefont {T.}~\bibnamefont
			{Coopmans}}, \bibinfo {author} {\bibfnamefont {L.}~\bibnamefont {Wubben}},
		\bibinfo {author} {\bibfnamefont {F.}~\bibnamefont {Rozp{\k{e}}dek}},
		\bibinfo {author} {\bibfnamefont {M.}~\bibnamefont {Pompili}}, \bibinfo
		{author} {\bibfnamefont {A.}~\bibnamefont {Stolk}}, \bibinfo {author}
		{\bibfnamefont {P.}~\bibnamefont {Pawe{\l}czak}}, \bibinfo {author}
		{\bibfnamefont {R.}~\bibnamefont {Knegjens}}, \bibinfo {author}
		{\bibfnamefont {R.}~\bibnamefont {Hanson}},  \emph {et~al.},\ }\href@noop {}
	{\bibfield  {journal} {\bibinfo  {journal} {arXiv preprint arXiv:1903.09778}\
		} (\bibinfo {year} {2019})}\BibitemShut {NoStop}%
	\bibitem [{\citenamefont {Matsuo}\ \emph {et~al.}(2019)\citenamefont {Matsuo},
		\citenamefont {Durand},\ and\ \citenamefont {Van~Meter}}]{matsuo2019quantum}%
	\BibitemOpen
	\bibfield  {author} {\bibinfo {author} {\bibfnamefont {T.}~\bibnamefont
			{Matsuo}}, \bibinfo {author} {\bibfnamefont {C.}~\bibnamefont {Durand}}, \
		and\ \bibinfo {author} {\bibfnamefont {R.}~\bibnamefont {Van~Meter}},\
	}\href@noop {} {\bibfield  {journal} {\bibinfo  {journal} {arXiv preprint
				arXiv:1904.08605}\ } (\bibinfo {year} {2019})}\BibitemShut {NoStop}%
	\bibitem [{\citenamefont {Jones}\ \emph {et~al.}(2016)\citenamefont {Jones},
		\citenamefont {Kim}, \citenamefont {Rakher}, \citenamefont {Kwiat},\ and\
		\citenamefont {Ladd}}]{jones16:comm-pro}%
	\BibitemOpen
	\bibfield  {author} {\bibinfo {author} {\bibfnamefont {C.}~\bibnamefont
			{Jones}}, \bibinfo {author} {\bibfnamefont {D.}~\bibnamefont {Kim}}, \bibinfo
		{author} {\bibfnamefont {M.~T.}\ \bibnamefont {Rakher}}, \bibinfo {author}
		{\bibfnamefont {P.~G.}\ \bibnamefont {Kwiat}}, \ and\ \bibinfo {author}
		{\bibfnamefont {T.~D.}\ \bibnamefont {Ladd}},\ }\href
	{http://stacks.iop.org/1367-2630/18/i=8/a=083015} {\bibfield  {journal}
		{\bibinfo  {journal} {New Journal of Physics}\ }\textbf {\bibinfo {volume}
			{18}},\ \bibinfo {pages} {083015} (\bibinfo {year} {2016})}\BibitemShut
	{NoStop}%
	\bibitem [{\citenamefont {Humphreys}\ \emph {et~al.}(2018)\citenamefont
		{Humphreys}, \citenamefont {Kalb}, \citenamefont {Morits}, \citenamefont
		{Schouten}, \citenamefont {Vermeulen}, \citenamefont {Twitchen},
		\citenamefont {Markham},\ and\ \citenamefont
		{Hanson}}]{humphreys2018deterministic}%
	\BibitemOpen
	\bibfield  {author} {\bibinfo {author} {\bibfnamefont {P.~C.}\ \bibnamefont
			{Humphreys}}, \bibinfo {author} {\bibfnamefont {N.}~\bibnamefont {Kalb}},
		\bibinfo {author} {\bibfnamefont {J.~P.}\ \bibnamefont {Morits}}, \bibinfo
		{author} {\bibfnamefont {R.~N.}\ \bibnamefont {Schouten}}, \bibinfo {author}
		{\bibfnamefont {R.~F.}\ \bibnamefont {Vermeulen}}, \bibinfo {author}
		{\bibfnamefont {D.~J.}\ \bibnamefont {Twitchen}}, \bibinfo {author}
		{\bibfnamefont {M.}~\bibnamefont {Markham}}, \ and\ \bibinfo {author}
		{\bibfnamefont {R.}~\bibnamefont {Hanson}},\ }\href@noop {} {\bibfield
		{journal} {\bibinfo  {journal} {Nature}\ }\textbf {\bibinfo {volume} {558}},\
		\bibinfo {pages} {268} (\bibinfo {year} {2018})}\BibitemShut {NoStop}%
	\bibitem [{\citenamefont {Krutyanskiy}\ \emph {et~al.}(2019)\citenamefont
		{Krutyanskiy}, \citenamefont {Meraner}, \citenamefont {Schupp}, \citenamefont
		{Krcmarsky}, \citenamefont {Hainzer},\ and\ \citenamefont
		{Lanyon}}]{krutyanskiy2019light}%
	\BibitemOpen
	\bibfield  {author} {\bibinfo {author} {\bibfnamefont {V.}~\bibnamefont
			{Krutyanskiy}}, \bibinfo {author} {\bibfnamefont {M.}~\bibnamefont
			{Meraner}}, \bibinfo {author} {\bibfnamefont {J.}~\bibnamefont {Schupp}},
		\bibinfo {author} {\bibfnamefont {V.}~\bibnamefont {Krcmarsky}}, \bibinfo
		{author} {\bibfnamefont {H.}~\bibnamefont {Hainzer}}, \ and\ \bibinfo
		{author} {\bibfnamefont {B.}~\bibnamefont {Lanyon}},\ }\href@noop {}
	{\bibfield  {journal} {\bibinfo  {journal} {npj Quantum Information}\
		}\textbf {\bibinfo {volume} {5}},\ \bibinfo {pages} {72} (\bibinfo {year}
		{2019})}\BibitemShut {NoStop}%
	\bibitem [{\citenamefont {Yuan}\ \emph {et~al.}(2008)\citenamefont {Yuan},
		\citenamefont {Chen}, \citenamefont {Zhao}, \citenamefont {Chen},
		\citenamefont {Schmiedmayer},\ and\ \citenamefont
		{Pan}}]{yuan2008experimental}%
	\BibitemOpen
	\bibfield  {author} {\bibinfo {author} {\bibfnamefont {Z.-S.}\ \bibnamefont
			{Yuan}}, \bibinfo {author} {\bibfnamefont {Y.-A.}\ \bibnamefont {Chen}},
		\bibinfo {author} {\bibfnamefont {B.}~\bibnamefont {Zhao}}, \bibinfo {author}
		{\bibfnamefont {S.}~\bibnamefont {Chen}}, \bibinfo {author} {\bibfnamefont
			{J.}~\bibnamefont {Schmiedmayer}}, \ and\ \bibinfo {author} {\bibfnamefont
			{J.-W.}\ \bibnamefont {Pan}},\ }\href@noop {} {\bibfield  {journal} {\bibinfo
			{journal} {Nature}\ }\textbf {\bibinfo {volume} {454}},\ \bibinfo {pages}
		{1098} (\bibinfo {year} {2008})}\BibitemShut {NoStop}%
	\bibitem [{\citenamefont {Hasegawa}\ \emph {et~al.}(2019)\citenamefont
		{Hasegawa}, \citenamefont {Ikuta}, \citenamefont {Matsuda}, \citenamefont
		{Tamaki}, \citenamefont {Lo}, \citenamefont {Yamamoto}, \citenamefont
		{Azuma},\ and\ \citenamefont {Imoto}}]{hasegawa2019experimental}%
	\BibitemOpen
	\bibfield  {author} {\bibinfo {author} {\bibfnamefont {Y.}~\bibnamefont
			{Hasegawa}}, \bibinfo {author} {\bibfnamefont {R.}~\bibnamefont {Ikuta}},
		\bibinfo {author} {\bibfnamefont {N.}~\bibnamefont {Matsuda}}, \bibinfo
		{author} {\bibfnamefont {K.}~\bibnamefont {Tamaki}}, \bibinfo {author}
		{\bibfnamefont {H.-K.}\ \bibnamefont {Lo}}, \bibinfo {author} {\bibfnamefont
			{T.}~\bibnamefont {Yamamoto}}, \bibinfo {author} {\bibfnamefont
			{K.}~\bibnamefont {Azuma}}, \ and\ \bibinfo {author} {\bibfnamefont
			{N.}~\bibnamefont {Imoto}},\ }\href@noop {} {\bibfield  {journal} {\bibinfo
			{journal} {Nature Communications}\ }\textbf {\bibinfo {volume} {10}},\
		\bibinfo {pages} {378} (\bibinfo {year} {2019})}\BibitemShut {NoStop}%
	\bibitem [{\citenamefont {Li}\ \emph {et~al.}(2019)\citenamefont {Li},
		\citenamefont {Zhang}, \citenamefont {Yin}, \citenamefont {Liu},
		\citenamefont {Hu}, \citenamefont {Fang}, \citenamefont {Fei}, \citenamefont
		{Jiang}, \citenamefont {Zhang}, \citenamefont {Li} \emph
		{et~al.}}]{li2019experimental}%
	\BibitemOpen
	\bibfield  {author} {\bibinfo {author} {\bibfnamefont {Z.-D.}\ \bibnamefont
			{Li}}, \bibinfo {author} {\bibfnamefont {R.}~\bibnamefont {Zhang}}, \bibinfo
		{author} {\bibfnamefont {X.-F.}\ \bibnamefont {Yin}}, \bibinfo {author}
		{\bibfnamefont {L.-Z.}\ \bibnamefont {Liu}}, \bibinfo {author} {\bibfnamefont
			{Y.}~\bibnamefont {Hu}}, \bibinfo {author} {\bibfnamefont {Y.-Q.}\
			\bibnamefont {Fang}}, \bibinfo {author} {\bibfnamefont {Y.-Y.}\ \bibnamefont
			{Fei}}, \bibinfo {author} {\bibfnamefont {X.}~\bibnamefont {Jiang}}, \bibinfo
		{author} {\bibfnamefont {J.}~\bibnamefont {Zhang}}, \bibinfo {author}
		{\bibfnamefont {L.}~\bibnamefont {Li}},  \emph {et~al.},\ }\href@noop {}
	{\bibfield  {journal} {\bibinfo  {journal} {Nature Photonics}\ }\textbf
		{\bibinfo {volume} {13}},\ \bibinfo {pages} {644} (\bibinfo {year}
		{2019})}\BibitemShut {NoStop}%
	\bibitem [{\citenamefont {Hensen}\ \emph {et~al.}(2015)\citenamefont {Hensen},
		\citenamefont {Bernien}, \citenamefont {Dr{\'e}au}, \citenamefont {Reiserer},
		\citenamefont {Kalb}, \citenamefont {Blok}, \citenamefont {Ruitenberg},
		\citenamefont {Vermeulen}, \citenamefont {Schouten}, \citenamefont
		{Abell{\'a}n} \emph {et~al.}}]{hensen2015loophole}%
	\BibitemOpen
	\bibfield  {author} {\bibinfo {author} {\bibfnamefont {B.}~\bibnamefont
			{Hensen}}, \bibinfo {author} {\bibfnamefont {H.}~\bibnamefont {Bernien}},
		\bibinfo {author} {\bibfnamefont {A.~E.}\ \bibnamefont {Dr{\'e}au}}, \bibinfo
		{author} {\bibfnamefont {A.}~\bibnamefont {Reiserer}}, \bibinfo {author}
		{\bibfnamefont {N.}~\bibnamefont {Kalb}}, \bibinfo {author} {\bibfnamefont
			{M.~S.}\ \bibnamefont {Blok}}, \bibinfo {author} {\bibfnamefont
			{J.}~\bibnamefont {Ruitenberg}}, \bibinfo {author} {\bibfnamefont {R.~F.}\
			\bibnamefont {Vermeulen}}, \bibinfo {author} {\bibfnamefont {R.~N.}\
			\bibnamefont {Schouten}}, \bibinfo {author} {\bibfnamefont {C.}~\bibnamefont
			{Abell{\'a}n}},  \emph {et~al.},\ }\href@noop {} {\bibfield  {journal}
		{\bibinfo  {journal} {Nature}\ }\textbf {\bibinfo {volume} {526}},\ \bibinfo
		{pages} {682} (\bibinfo {year} {2015})}\BibitemShut {NoStop}%
	\bibitem [{\citenamefont {Rozp{\k{e}}dek}\ \emph {et~al.}(2019)\citenamefont
		{Rozp{\k{e}}dek}, \citenamefont {Yehia}, \citenamefont {Goodenough},
		\citenamefont {Ruf}, \citenamefont {Humphreys}, \citenamefont {Hanson},
		\citenamefont {Wehner},\ and\ \citenamefont {Elkouss}}]{rozpkedek2019near}%
	\BibitemOpen
	\bibfield  {author} {\bibinfo {author} {\bibfnamefont {F.}~\bibnamefont
			{Rozp{\k{e}}dek}}, \bibinfo {author} {\bibfnamefont {R.}~\bibnamefont
			{Yehia}}, \bibinfo {author} {\bibfnamefont {K.}~\bibnamefont {Goodenough}},
		\bibinfo {author} {\bibfnamefont {M.}~\bibnamefont {Ruf}}, \bibinfo {author}
		{\bibfnamefont {P.~C.}\ \bibnamefont {Humphreys}}, \bibinfo {author}
		{\bibfnamefont {R.}~\bibnamefont {Hanson}}, \bibinfo {author} {\bibfnamefont
			{S.}~\bibnamefont {Wehner}}, \ and\ \bibinfo {author} {\bibfnamefont
			{D.}~\bibnamefont {Elkouss}},\ }\href@noop {} {\bibfield  {journal} {\bibinfo
			{journal} {Physical Review A}\ }\textbf {\bibinfo {volume} {99}},\ \bibinfo
		{pages} {052330} (\bibinfo {year} {2019})}\BibitemShut {NoStop}%
	\bibitem [{\citenamefont {Kumar}\ \emph {et~al.}(2019)\citenamefont {Kumar},
		\citenamefont {Lauk},\ and\ \citenamefont {Simon}}]{Kumar_2019}%
	\BibitemOpen
	\bibfield  {author} {\bibinfo {author} {\bibfnamefont {S.}~\bibnamefont
			{Kumar}}, \bibinfo {author} {\bibfnamefont {N.}~\bibnamefont {Lauk}}, \ and\
		\bibinfo {author} {\bibfnamefont {C.}~\bibnamefont {Simon}},\ }\href
	{\doibase 10.1088/2058-9565/ab2c87} {\bibfield  {journal} {\bibinfo
			{journal} {Quantum Science and Technology}\ }\textbf {\bibinfo {volume}
			{4}},\ \bibinfo {pages} {045003} (\bibinfo {year} {2019})}\BibitemShut
	{NoStop}%
	\bibitem [{\citenamefont {Zukowski}\ \emph {et~al.}(1993)\citenamefont
		{Zukowski}, \citenamefont {Zeilinger}, \citenamefont {Horne},\ and\
		\citenamefont {Ekert}}]{zukowski1993event}%
	\BibitemOpen
	\bibfield  {author} {\bibinfo {author} {\bibfnamefont {M.}~\bibnamefont
			{Zukowski}}, \bibinfo {author} {\bibfnamefont {A.}~\bibnamefont {Zeilinger}},
		\bibinfo {author} {\bibfnamefont {M.~A.}\ \bibnamefont {Horne}}, \ and\
		\bibinfo {author} {\bibfnamefont {A.~K.}\ \bibnamefont {Ekert}},\ }\href@noop
	{} {\bibfield  {journal} {\bibinfo  {journal} {Physical Review Letters}\
		}\textbf {\bibinfo {volume} {71}},\ \bibinfo {pages} {4287} (\bibinfo {year}
		{1993})}\BibitemShut {NoStop}%
	\bibitem [{\citenamefont {Pan}\ \emph {et~al.}(1998)\citenamefont {Pan},
		\citenamefont {Bouwmeester}, \citenamefont {Weinfurter},\ and\ \citenamefont
		{Zeilinger}}]{EntanglementSwappingOriginal}%
	\BibitemOpen
	\bibfield  {author} {\bibinfo {author} {\bibfnamefont {J.-W.}\ \bibnamefont
			{Pan}}, \bibinfo {author} {\bibfnamefont {D.}~\bibnamefont {Bouwmeester}},
		\bibinfo {author} {\bibfnamefont {H.}~\bibnamefont {Weinfurter}}, \ and\
		\bibinfo {author} {\bibfnamefont {A.}~\bibnamefont {Zeilinger}},\ }\href
	{\doibase 10.1103/PhysRevLett.80.3891} {\bibfield  {journal} {\bibinfo
			{journal} {Phys. Rev. Lett.}\ }\textbf {\bibinfo {volume} {80}},\ \bibinfo
		{pages} {3891} (\bibinfo {year} {1998})}\BibitemShut {NoStop}%
	\bibitem [{\citenamefont {Raussendorf}\ and\ \citenamefont
		{Briegel}(2001)}]{MBQC}%
	\BibitemOpen
	\bibfield  {author} {\bibinfo {author} {\bibfnamefont {R.}~\bibnamefont
			{Raussendorf}}\ and\ \bibinfo {author} {\bibfnamefont {H.~J.}\ \bibnamefont
			{Briegel}},\ }\href {\doibase 10.1103/PhysRevLett.86.5188} {\bibfield
		{journal} {\bibinfo  {journal} {Phys. Rev. Lett.}\ }\textbf {\bibinfo
			{volume} {86}},\ \bibinfo {pages} {5188} (\bibinfo {year}
		{2001})}\BibitemShut {NoStop}%
	\bibitem [{\citenamefont {Matsuo}\ \emph {et~al.}(2018)\citenamefont {Matsuo},
		\citenamefont {Satoh}, \citenamefont {Nagayama},\ and\ \citenamefont
		{Van~Meter}}]{MQNC}%
	\BibitemOpen
	\bibfield  {author} {\bibinfo {author} {\bibfnamefont {T.}~\bibnamefont
			{Matsuo}}, \bibinfo {author} {\bibfnamefont {T.}~\bibnamefont {Satoh}},
		\bibinfo {author} {\bibfnamefont {S.}~\bibnamefont {Nagayama}}, \ and\
		\bibinfo {author} {\bibfnamefont {R.}~\bibnamefont {Van~Meter}},\ }\href
	{\doibase 10.1103/PhysRevA.97.062328} {\bibfield  {journal} {\bibinfo
			{journal} {Phys. Rev. A}\ }\textbf {\bibinfo {volume} {97}},\ \bibinfo
		{pages} {062328} (\bibinfo {year} {2018})}\BibitemShut {NoStop}%
	\bibitem [{\citenamefont {Clauser}\ \emph {et~al.}(1969)\citenamefont
		{Clauser}, \citenamefont {Horne}, \citenamefont {Shimony},\ and\
		\citenamefont {Holt}}]{CHSHOriginalPaper}%
	\BibitemOpen
	\bibfield  {author} {\bibinfo {author} {\bibfnamefont {J.~F.}\ \bibnamefont
			{Clauser}}, \bibinfo {author} {\bibfnamefont {M.~A.}\ \bibnamefont {Horne}},
		\bibinfo {author} {\bibfnamefont {A.}~\bibnamefont {Shimony}}, \ and\
		\bibinfo {author} {\bibfnamefont {R.~A.}\ \bibnamefont {Holt}},\ }\href
	{\doibase 10.1103/PhysRevLett.23.880} {\bibfield  {journal} {\bibinfo
			{journal} {Phys. Rev. Lett.}\ }\textbf {\bibinfo {volume} {23}},\ \bibinfo
		{pages} {880} (\bibinfo {year} {1969})}\BibitemShut {NoStop}%
	\bibitem [{\citenamefont {Van~Meter}\ \emph {et~al.}(2008)\citenamefont
		{Van~Meter}, \citenamefont {Munro}, \citenamefont {Nemoto},\ and\
		\citenamefont {Itoh}}]{van-meter07:_distr_arith_jetc}%
	\BibitemOpen
	\bibfield  {author} {\bibinfo {author} {\bibfnamefont {R.}~\bibnamefont
			{Van~Meter}}, \bibinfo {author} {\bibfnamefont {W.~J.}\ \bibnamefont
			{Munro}}, \bibinfo {author} {\bibfnamefont {K.}~\bibnamefont {Nemoto}}, \
		and\ \bibinfo {author} {\bibfnamefont {K.~M.}\ \bibnamefont {Itoh}},\ }\href
	{\doibase 10.1145/1324177.1324179} {\bibfield  {journal} {\bibinfo  {journal}
			{J. Emerg. Technol. Comput. Syst.}\ }\textbf {\bibinfo {volume} {3}},\
		\bibinfo {pages} {2:1} (\bibinfo {year} {2008})}\BibitemShut {NoStop}%
	\bibitem [{\citenamefont {Nishio}\ \emph {et~al.}(2019)\citenamefont {Nishio},
		\citenamefont {Pan}, \citenamefont {Satoh}, \citenamefont {Amano},\ and\
		\citenamefont {Van~Meter}}]{1903.10963}%
	\BibitemOpen
	\bibfield  {author} {\bibinfo {author} {\bibfnamefont {S.}~\bibnamefont
			{Nishio}}, \bibinfo {author} {\bibfnamefont {Y.}~\bibnamefont {Pan}},
		\bibinfo {author} {\bibfnamefont {T.}~\bibnamefont {Satoh}}, \bibinfo
		{author} {\bibfnamefont {H.}~\bibnamefont {Amano}}, \ and\ \bibinfo {author}
		{\bibfnamefont {R.}~\bibnamefont {Van~Meter}},\ }\href@noop {} {\enquote
		{\bibinfo {title} {{E}xtracting {S}uccess from {IBM's} 20-{Q}ubit {M}achines
				{U}sing {E}rror-{A}ware {C}ompilation},}\ } (\bibinfo {year} {2019}),\
	\Eprint {http://arxiv.org/abs/arXiv:1903.10963v1} {arXiv:1903.10963v1}
	\BibitemShut {NoStop}%
	\bibitem [{\citenamefont {Bennett}\ \emph {et~al.}(1993)\citenamefont
		{Bennett}, \citenamefont {Brassard}, \citenamefont {Cr\'epeau}, \citenamefont
		{Jozsa}, \citenamefont {Peres},\ and\ \citenamefont
		{Wootters}}]{bennett:teleportation}%
	\BibitemOpen
	\bibfield  {author} {\bibinfo {author} {\bibfnamefont {C.~H.}\ \bibnamefont
			{Bennett}}, \bibinfo {author} {\bibfnamefont {G.}~\bibnamefont {Brassard}},
		\bibinfo {author} {\bibfnamefont {C.}~\bibnamefont {Cr\'epeau}}, \bibinfo
		{author} {\bibfnamefont {R.}~\bibnamefont {Jozsa}}, \bibinfo {author}
		{\bibfnamefont {A.}~\bibnamefont {Peres}}, \ and\ \bibinfo {author}
		{\bibfnamefont {W.~K.}\ \bibnamefont {Wootters}},\ }\href {\doibase
		10.1103/PhysRevLett.70.1895} {\bibfield  {journal} {\bibinfo  {journal}
			{Phys. Rev. Lett.}\ }\textbf {\bibinfo {volume} {70}},\ \bibinfo {pages}
		{1895} (\bibinfo {year} {1993})}\BibitemShut {NoStop}%
	\bibitem [{\citenamefont {Eisert}\ \emph {et~al.}(2000)\citenamefont {Eisert},
		\citenamefont {Jacobs}, \citenamefont {Papadopoulos},\ and\ \citenamefont
		{Plenio}}]{eisert2000oli}%
	\BibitemOpen
	\bibfield  {author} {\bibinfo {author} {\bibfnamefont {J.}~\bibnamefont
			{Eisert}}, \bibinfo {author} {\bibfnamefont {K.}~\bibnamefont {Jacobs}},
		\bibinfo {author} {\bibfnamefont {P.}~\bibnamefont {Papadopoulos}}, \ and\
		\bibinfo {author} {\bibfnamefont {M.}~\bibnamefont {Plenio}},\ }\href@noop {}
	{\bibfield  {journal} {\bibinfo  {journal} {Physical Review A}\ }\textbf
		{\bibinfo {volume} {62}},\ \bibinfo {pages} {52317} (\bibinfo {year}
		{2000})}\BibitemShut {NoStop}%
	\bibitem [{\citenamefont {Gottesman}\ and\ \citenamefont
		{Chuang}(1999)}]{gottsman99:universal_teleport}%
	\BibitemOpen
	\bibfield  {author} {\bibinfo {author} {\bibfnamefont {D.}~\bibnamefont
			{Gottesman}}\ and\ \bibinfo {author} {\bibfnamefont {I.~L.}\ \bibnamefont
			{Chuang}},\ }\href@noop {} {\bibfield  {journal} {\bibinfo  {journal}
			{Nature}\ }\textbf {\bibinfo {volume} {402}},\ \bibinfo {pages} {390}
		(\bibinfo {year} {1999})}\BibitemShut {NoStop}%
	\bibitem [{\citenamefont {Raussendorf}\ \emph {et~al.}(2003)\citenamefont
		{Raussendorf}, \citenamefont {Browne},\ and\ \citenamefont
		{Briegel}}]{MBQC-2}%
	\BibitemOpen
	\bibfield  {author} {\bibinfo {author} {\bibfnamefont {R.}~\bibnamefont
			{Raussendorf}}, \bibinfo {author} {\bibfnamefont {D.~E.}\ \bibnamefont
			{Browne}}, \ and\ \bibinfo {author} {\bibfnamefont {H.~J.}\ \bibnamefont
			{Briegel}},\ }\href {\doibase 10.1103/PhysRevA.68.022312} {\bibfield
		{journal} {\bibinfo  {journal} {Phys. Rev. A}\ }\textbf {\bibinfo {volume}
			{68}},\ \bibinfo {pages} {022312} (\bibinfo {year} {2003})}\BibitemShut
	{NoStop}%
	\bibitem [{\citenamefont {Aparicio}\ and\ \citenamefont {Van{
			}Meter}(2011)}]{aparicio11:repeater-muxing}%
	\BibitemOpen
	\bibfield  {author} {\bibinfo {author} {\bibfnamefont {L.}~\bibnamefont
			{Aparicio}}\ and\ \bibinfo {author} {\bibfnamefont {R.}~\bibnamefont {Van{
				}Meter}},\ }in\ \href@noop {} {\emph {\bibinfo {booktitle} {Proc. SPIE}}},\
	Vol.\ \bibinfo {volume} {8163}\ (\bibinfo {year} {2011})\ p.\ \bibinfo
	{pages} {816308}\BibitemShut {NoStop}%
	\bibitem [{\citenamefont {Briegel}\ and\ \citenamefont
		{Raussendorf}(2001)}]{briegel2001persistent}%
	\BibitemOpen
	\bibfield  {author} {\bibinfo {author} {\bibfnamefont {H.~J.}\ \bibnamefont
			{Briegel}}\ and\ \bibinfo {author} {\bibfnamefont {R.}~\bibnamefont
			{Raussendorf}},\ }\href@noop {} {\bibfield  {journal} {\bibinfo  {journal}
			{Physical Review Letters}\ }\textbf {\bibinfo {volume} {86}},\ \bibinfo
		{pages} {910} (\bibinfo {year} {2001})}\BibitemShut {NoStop}%
	\bibitem [{\citenamefont {Hajdu{\v{s}}ek}\ and\ \citenamefont
		{Vedral}(2010)}]{hajduvsek2010entanglement}%
	\BibitemOpen
	\bibfield  {author} {\bibinfo {author} {\bibfnamefont {M.}~\bibnamefont
			{Hajdu{\v{s}}ek}}\ and\ \bibinfo {author} {\bibfnamefont {V.}~\bibnamefont
			{Vedral}},\ }\href@noop {} {\bibfield  {journal} {\bibinfo  {journal} {New
				Journal of Physics}\ }\textbf {\bibinfo {volume} {12}},\ \bibinfo {pages}
		{053015} (\bibinfo {year} {2010})}\BibitemShut {NoStop}%
	\bibitem [{\citenamefont {Hajdu{\v{s}}ek}\ and\ \citenamefont
		{Murao}(2013)}]{hajduvsek2013direct}%
	\BibitemOpen
	\bibfield  {author} {\bibinfo {author} {\bibfnamefont {M.}~\bibnamefont
			{Hajdu{\v{s}}ek}}\ and\ \bibinfo {author} {\bibfnamefont {M.}~\bibnamefont
			{Murao}},\ }\href@noop {} {\bibfield  {journal} {\bibinfo  {journal} {New
				Journal of Physics}\ }\textbf {\bibinfo {volume} {15}},\ \bibinfo {pages}
		{013039} (\bibinfo {year} {2013})}\BibitemShut {NoStop}%
	\bibitem [{\citenamefont {Van{ }Meter}\ \emph {et~al.}(2013)\citenamefont {Van{
			}Meter}, \citenamefont {Satoh}, \citenamefont {Ladd}, \citenamefont {Munro},\
		and\ \citenamefont {Nemoto}}]{van-meter:qDijkstra}%
	\BibitemOpen
	\bibfield  {author} {\bibinfo {author} {\bibfnamefont {R.}~\bibnamefont {Van{
				}Meter}}, \bibinfo {author} {\bibfnamefont {T.}~\bibnamefont {Satoh}},
		\bibinfo {author} {\bibfnamefont {T.~D.}\ \bibnamefont {Ladd}}, \bibinfo
		{author} {\bibfnamefont {W.~J.}\ \bibnamefont {Munro}}, \ and\ \bibinfo
		{author} {\bibfnamefont {K.}~\bibnamefont {Nemoto}},\ }\href {\doibase
		10.1007/s13119-013-0026-2} {\bibfield  {journal} {\bibinfo  {journal}
			{Networking Science}\ }\textbf {\bibinfo {volume} {3}},\ \bibinfo {pages}
		{82} (\bibinfo {year} {2013})}\BibitemShut {NoStop}%
	\bibitem [{\citenamefont {Caleffi}(2017)}]{caleffi17:routing}%
	\BibitemOpen
	\bibfield  {author} {\bibinfo {author} {\bibfnamefont {M.}~\bibnamefont
			{Caleffi}},\ }\href {\doibase 10.1109/ACCESS.2017.2763325} {\bibfield
		{journal} {\bibinfo  {journal} {IEEE Access}\ }\textbf {\bibinfo {volume}
			{5}},\ \bibinfo {pages} {22299} (\bibinfo {year} {2017})}\BibitemShut
	{NoStop}%
	\bibitem [{\citenamefont {Schoute}(2015)}]{schoute15:_shortcut_thesis}%
	\BibitemOpen
	\bibfield  {author} {\bibinfo {author} {\bibfnamefont {E.}~\bibnamefont
			{Schoute}},\ }\emph {\bibinfo {title} {Shortcuts to quantum network
			routing}},\ \href
	{https://repository.tudelft.nl/islandora/object/uuid:684bb201-64a2-461d-b603-1f47590a1341/?collection=research}
	{Master's thesis},\ \bibinfo  {school} {T.U. Delft} (\bibinfo {year}
	{2015})\BibitemShut {NoStop}%
	\bibitem [{\citenamefont {Pant}\ \emph {et~al.}(2019)\citenamefont {Pant},
		\citenamefont {Krovi}, \citenamefont {Towsley}, \citenamefont {Tassiulas},
		\citenamefont {Jiang}, \citenamefont {Basu}, \citenamefont {Englund},\ and\
		\citenamefont {Guha}}]{Pant2019}%
	\BibitemOpen
	\bibfield  {author} {\bibinfo {author} {\bibfnamefont {M.}~\bibnamefont
			{Pant}}, \bibinfo {author} {\bibfnamefont {H.}~\bibnamefont {Krovi}},
		\bibinfo {author} {\bibfnamefont {D.}~\bibnamefont {Towsley}}, \bibinfo
		{author} {\bibfnamefont {L.}~\bibnamefont {Tassiulas}}, \bibinfo {author}
		{\bibfnamefont {L.}~\bibnamefont {Jiang}}, \bibinfo {author} {\bibfnamefont
			{P.}~\bibnamefont {Basu}}, \bibinfo {author} {\bibfnamefont {D.}~\bibnamefont
			{Englund}}, \ and\ \bibinfo {author} {\bibfnamefont {S.}~\bibnamefont
			{Guha}},\ }\href {\doibase 10.1038/s41534-019-0139-x} {\bibfield  {journal}
		{\bibinfo  {journal} {npj Quantum Information}\ }\textbf {\bibinfo {volume}
			{5}},\ \bibinfo {pages} {25} (\bibinfo {year} {2019})}\BibitemShut {NoStop}%
	\bibitem [{\citenamefont {Gyongyosi}\ and\ \citenamefont
		{Imre}(2017)}]{gyongyosi2017entanglement}%
	\BibitemOpen
	\bibfield  {author} {\bibinfo {author} {\bibfnamefont {L.}~\bibnamefont
			{Gyongyosi}}\ and\ \bibinfo {author} {\bibfnamefont {S.}~\bibnamefont
			{Imre}},\ }\href@noop {} {\bibfield  {journal} {\bibinfo  {journal}
			{Scientific reports}\ }\textbf {\bibinfo {volume} {7}},\ \bibinfo {pages}
		{14255} (\bibinfo {year} {2017})}\BibitemShut {NoStop}%
	\bibitem [{\citenamefont {Behera}\ \emph {et~al.}(2019)\citenamefont {Behera},
		\citenamefont {Reza}, \citenamefont {Gupta},\ and\ \citenamefont
		{Panigrahi}}]{Behera2019}%
	\BibitemOpen
	\bibfield  {author} {\bibinfo {author} {\bibfnamefont {B.~K.}\ \bibnamefont
			{Behera}}, \bibinfo {author} {\bibfnamefont {T.}~\bibnamefont {Reza}},
		\bibinfo {author} {\bibfnamefont {A.}~\bibnamefont {Gupta}}, \ and\ \bibinfo
		{author} {\bibfnamefont {P.~K.}\ \bibnamefont {Panigrahi}},\ }\href {\doibase
		10.1007/s11128-019-2436-x} {\bibfield  {journal} {\bibinfo  {journal}
			{Quantum Information Processing}\ }\textbf {\bibinfo {volume} {18}},\
		\bibinfo {pages} {328} (\bibinfo {year} {2019})}\BibitemShut {NoStop}%
	\bibitem [{\citenamefont {Jacobson}(1988)}]{Jacobson:1988:CAC:52324.52356}%
	\BibitemOpen
	\bibfield  {author} {\bibinfo {author} {\bibfnamefont {V.}~\bibnamefont
			{Jacobson}},\ }in\ \href {\doibase 10.1145/52324.52356} {\emph {\bibinfo
			{booktitle} {Symposium proceedings on Communications architectures and
				protocols}}},\ \bibinfo {series and number} {SIGCOMM '88}\ (\bibinfo
	{publisher} {ACM},\ \bibinfo {address} {New York, NY, USA},\ \bibinfo {year}
	{1988})\ pp.\ \bibinfo {pages} {314--329}\BibitemShut {NoStop}%
	\bibitem [{\citenamefont {Ahlswede}\ \emph {et~al.}(2000)\citenamefont
		{Ahlswede}, \citenamefont {Cai}, \citenamefont {Li},\ and\ \citenamefont
		{Yeung}}]{ahlswede2000nif}%
	\BibitemOpen
	\bibfield  {author} {\bibinfo {author} {\bibfnamefont {R.}~\bibnamefont
			{Ahlswede}}, \bibinfo {author} {\bibfnamefont {N.}~\bibnamefont {Cai}},
		\bibinfo {author} {\bibfnamefont {S.}~\bibnamefont {Li}}, \ and\ \bibinfo
		{author} {\bibfnamefont {R.}~\bibnamefont {Yeung}},\ }\href@noop {}
	{\bibfield  {journal} {\bibinfo  {journal} {Information Theory, IEEE
				Transactions on}\ }\textbf {\bibinfo {volume} {46}},\ \bibinfo {pages} {1204}
		(\bibinfo {year} {2000})}\BibitemShut {NoStop}%
	\bibitem [{\citenamefont {Hayashi}\ \emph {et~al.}(2007)\citenamefont
		{Hayashi}, \citenamefont {Iwama}, \citenamefont {Nishimura}, \citenamefont
		{Raymond},\ and\ \citenamefont {Yamashita}}]{hayashi2006qnc}%
	\BibitemOpen
	\bibfield  {author} {\bibinfo {author} {\bibfnamefont {M.}~\bibnamefont
			{Hayashi}}, \bibinfo {author} {\bibfnamefont {K.}~\bibnamefont {Iwama}},
		\bibinfo {author} {\bibfnamefont {H.}~\bibnamefont {Nishimura}}, \bibinfo
		{author} {\bibfnamefont {R.}~\bibnamefont {Raymond}}, \ and\ \bibinfo
		{author} {\bibfnamefont {S.}~\bibnamefont {Yamashita}},\ }in\ \href@noop {}
	{\emph {\bibinfo {booktitle} {STACS 2007}}},\ \bibinfo {editor} {edited by\
		\bibinfo {editor} {\bibfnamefont {W.}~\bibnamefont {Thomas}}\ and\ \bibinfo
		{editor} {\bibfnamefont {P.}~\bibnamefont {Weil}}}\ (\bibinfo  {publisher}
	{Springer Berlin Heidelberg},\ \bibinfo {address} {Berlin, Heidelberg},\
	\bibinfo {year} {2007})\ pp.\ \bibinfo {pages} {610--621}\BibitemShut
	{NoStop}%
	\bibitem [{\citenamefont {Iwama}\ \emph {et~al.}(2006)\citenamefont {Iwama},
		\citenamefont {Nishimura}, \citenamefont {Raymond},\ and\ \citenamefont
		{Yamashita}}]{iwama2006qnc}%
	\BibitemOpen
	\bibfield  {author} {\bibinfo {author} {\bibfnamefont {K.}~\bibnamefont
			{Iwama}}, \bibinfo {author} {\bibfnamefont {H.}~\bibnamefont {Nishimura}},
		\bibinfo {author} {\bibfnamefont {R.}~\bibnamefont {Raymond}}, \ and\
		\bibinfo {author} {\bibfnamefont {S.}~\bibnamefont {Yamashita}},\ }\href@noop
	{} {\bibfield  {journal} {\bibinfo  {journal} {Arxiv preprint
				quant-ph/0611039}\ } (\bibinfo {year} {2006})}\BibitemShut {NoStop}%
	\bibitem [{\citenamefont {Leung}\ \emph {et~al.}(2010)\citenamefont {Leung},
		\citenamefont {Oppenheim},\ and\ \citenamefont {Winter}}]{leung:1840528}%
	\BibitemOpen
	\bibfield  {author} {\bibinfo {author} {\bibfnamefont {D.}~\bibnamefont
			{Leung}}, \bibinfo {author} {\bibfnamefont {J.}~\bibnamefont {Oppenheim}}, \
		and\ \bibinfo {author} {\bibfnamefont {A.}~\bibnamefont {Winter}},\ }\href
	{\doibase http://dx.doi.org/10.1109/TIT.2010.2048442} {\bibfield  {journal}
		{\bibinfo  {journal} {IEEE Trans. Inf. Theor.}\ }\textbf {\bibinfo {volume}
			{56}},\ \bibinfo {pages} {3478} (\bibinfo {year} {2010})}\BibitemShut
	{NoStop}%
	\bibitem [{\citenamefont {Satoh}\ \emph {et~al.}(2016)\citenamefont {Satoh},
		\citenamefont {Ishizaki}, \citenamefont {Nagayama},\ and\ \citenamefont
		{Van~Meter}}]{satoh:PhysRevA.93.032302}%
	\BibitemOpen
	\bibfield  {author} {\bibinfo {author} {\bibfnamefont {T.}~\bibnamefont
			{Satoh}}, \bibinfo {author} {\bibfnamefont {K.}~\bibnamefont {Ishizaki}},
		\bibinfo {author} {\bibfnamefont {S.}~\bibnamefont {Nagayama}}, \ and\
		\bibinfo {author} {\bibfnamefont {R.}~\bibnamefont {Van~Meter}},\ }\href
	{\doibase 10.1103/PhysRevA.93.032302} {\bibfield  {journal} {\bibinfo
			{journal} {Phys. Rev. A}\ }\textbf {\bibinfo {volume} {93}},\ \bibinfo
		{pages} {032302} (\bibinfo {year} {2016})}\BibitemShut {NoStop}%
	\bibitem [{\citenamefont {Satoh}\ \emph {et~al.}(2012)\citenamefont {Satoh},
		\citenamefont {Le~Gall},\ and\ \citenamefont {Imai}}]{QNC}%
	\BibitemOpen
	\bibfield  {author} {\bibinfo {author} {\bibfnamefont {T.}~\bibnamefont
			{Satoh}}, \bibinfo {author} {\bibfnamefont {F.~m.~c.}\ \bibnamefont
			{Le~Gall}}, \ and\ \bibinfo {author} {\bibfnamefont {H.}~\bibnamefont
			{Imai}},\ }\href {\doibase 10.1103/PhysRevA.86.032331} {\bibfield  {journal}
		{\bibinfo  {journal} {Phys. Rev. A}\ }\textbf {\bibinfo {volume} {86}},\
		\bibinfo {pages} {032331} (\bibinfo {year} {2012})}\BibitemShut {NoStop}%
	\bibitem [{\citenamefont {{Akibue}}\ and\ \citenamefont
		{{Murao}}(2016)}]{akibue:7556338}%
	\BibitemOpen
	\bibfield  {author} {\bibinfo {author} {\bibfnamefont {S.}~\bibnamefont
			{{Akibue}}}\ and\ \bibinfo {author} {\bibfnamefont {M.}~\bibnamefont
			{{Murao}}},\ }\href {\doibase 10.1109/TIT.2016.2604382} {\bibfield  {journal}
		{\bibinfo  {journal} {IEEE Transactions on Information Theory}\ }\textbf
		{\bibinfo {volume} {62}},\ \bibinfo {pages} {6620} (\bibinfo {year}
		{2016})}\BibitemShut {NoStop}%
	\bibitem [{\citenamefont {Abraham}\ \emph {et~al.}(2019)\citenamefont
		{Abraham}, \citenamefont {Akhalwaya}, \citenamefont {Aleksandrowicz},
		\citenamefont {Alexander}, \citenamefont {Alexandrowics}, \citenamefont
		{Arbel}, \citenamefont {Asfaw}, \citenamefont {Azaustre}, \citenamefont
		{Barkoutsos}, \citenamefont {Barron}, \citenamefont {Bello}, \citenamefont
		{Ben-Haim}, \citenamefont {Bishop}, \citenamefont {Bosch}, \citenamefont
		{Bucher}, \citenamefont {CZ}, \citenamefont {Cabrera}, \citenamefont
		{Calpin}, \citenamefont {Capelluto}, \citenamefont {Carballo}, \citenamefont
		{Chen}, \citenamefont {Chen}, \citenamefont {Chen}, \citenamefont {Chow},
		\citenamefont {Claus}, \citenamefont {Cross}, \citenamefont {Cross},
		\citenamefont {Cruz-Benito}, \citenamefont {Cryoris}, \citenamefont {Culver},
		\citenamefont {C{\'o}rcoles-Gonzales}, \citenamefont {Dague}, \citenamefont
		{Dartiailh}, \citenamefont {Davila}, \citenamefont {Ding}, \citenamefont
		{Dumitrescu}, \citenamefont {Dumon}, \citenamefont {Duran}, \citenamefont
		{Eendebak}, \citenamefont {Egger}, \citenamefont {Everitt}, \citenamefont
		{Fern{\'a}ndez}, \citenamefont {Frisch}, \citenamefont {Fuhrer},
		\citenamefont {Gacon}, \citenamefont {Gadi}, \citenamefont {Gago},
		\citenamefont {Gambetta}, \citenamefont {Garcia}, \citenamefont {Garion},
		\citenamefont {Gawel-Kus}, \citenamefont {Gil}, \citenamefont
		{Gomez-Mosquera}, \citenamefont {de~la Puente~Gonz{\'a}lez}, \citenamefont
		{Greenberg}, \citenamefont {Gunnels}, \citenamefont {Haide}, \citenamefont
		{Hamamura}, \citenamefont {Havlicek}, \citenamefont {Hellmers}, \citenamefont
		{Herok}, \citenamefont {Horii}, \citenamefont {Howington}, \citenamefont
		{Hu}, \citenamefont {Hu}, \citenamefont {Imai}, \citenamefont {Imamichi},
		\citenamefont {Iten}, \citenamefont {Itoko}, \citenamefont {Javadi-Abhari},
		\citenamefont {Jessica}, \citenamefont {Johns}, \citenamefont {Kanazawa},
		\citenamefont {Karazeev}, \citenamefont {Kassebaum}, \citenamefont
		{Krishnan}, \citenamefont {Krsulich}, \citenamefont {Kus}, \citenamefont
		{LaRose}, \citenamefont {Lambert}, \citenamefont {Latone}, \citenamefont
		{Lawrence}, \citenamefont {Liu}, \citenamefont {Mac}, \citenamefont {Maeng},
		\citenamefont {Malyshev}, \citenamefont {Marecek}, \citenamefont {Marques},
		\citenamefont {Mathews}, \citenamefont {Matsuo}, \citenamefont {McClure},
		\citenamefont {McGarry}, \citenamefont {McKay}, \citenamefont {Meesala},
		\citenamefont {Mezzacapo}, \citenamefont {Midha}, \citenamefont {Minev},
		\citenamefont {Morales}, \citenamefont {Murali}, \citenamefont
		{M{\"u}ggenburg}, \citenamefont {Nadlinger}, \citenamefont {Nannicini},
		\citenamefont {Nation}, \citenamefont {Naveh}, \citenamefont
		{Nick-Singstock}, \citenamefont {Niroula}, \citenamefont {Norlen},
		\citenamefont {O'Riordan}, \citenamefont {Ollitrault}, \citenamefont {Oud},
		\citenamefont {Padilha}, \citenamefont {Paik}, \citenamefont {Perriello},
		\citenamefont {Phan}, \citenamefont {Pistoia}, \citenamefont
		{Pozas-iKerstjens}, \citenamefont {Prutyanov}, \citenamefont {P{\'e}rez},
		\citenamefont {Quintiii}, \citenamefont {Raymond}, \citenamefont {Redondo},
		\citenamefont {Reuter}, \citenamefont {Rodr{\'\i}guez}, \citenamefont {Ryu},
		\citenamefont {Sandberg}, \citenamefont {Sathaye}, \citenamefont {Schmitt},
		\citenamefont {Schnabel}, \citenamefont {Scholten}, \citenamefont {Schoute},
		\citenamefont {Sertage}, \citenamefont {Shi}, \citenamefont {Silva},
		\citenamefont {Siraichi}, \citenamefont {Sivarajah}, \citenamefont {Smolin},
		\citenamefont {Soeken}, \citenamefont {Steenken}, \citenamefont
		{Stypulkoski}, \citenamefont {Takahashi}, \citenamefont {Taylor},
		\citenamefont {Taylour}, \citenamefont {Thomas}, \citenamefont {Tillet},
		\citenamefont {Tod}, \citenamefont {de~la Torre}, \citenamefont {Trabing},
		\citenamefont {Treinish}, \citenamefont {TrishaPe}, \citenamefont {Turner},
		\citenamefont {Vaknin}, \citenamefont {Valcarce}, \citenamefont {Varchon},
		\citenamefont {Vogt-Lee}, \citenamefont {Vuillot}, \citenamefont {Weaver},
		\citenamefont {Wieczorek}, \citenamefont {Wildstrom}, \citenamefont {Wille},
		\citenamefont {Winston}, \citenamefont {Woehr}, \citenamefont {Woerner},
		\citenamefont {Woo}, \citenamefont {Wood}, \citenamefont {Wood},
		\citenamefont {Wood}, \citenamefont {Wootton}, \citenamefont {Yeralin},
		\citenamefont {Yu}, \citenamefont {Zdanski}, \citenamefont {Zoufalc},
		\citenamefont {anedumla}, \citenamefont {azulehner}, \citenamefont
		{bcamorrison}, \citenamefont {drholmie}, \citenamefont {fanizzamarco},
		\citenamefont {kanejess}, \citenamefont {klinvill}, \citenamefont {merav
			aharoni}, \citenamefont {ordmoj}, \citenamefont {tigerjack}, \citenamefont
		{yang.luh},\ and\ \citenamefont {yotamvakninibm}}]{Qiskit}%
	\BibitemOpen
	\bibfield  {author} {\bibinfo {author} {\bibfnamefont {H.}~\bibnamefont
			{Abraham}}, \bibinfo {author} {\bibfnamefont {I.~Y.}\ \bibnamefont
			{Akhalwaya}}, \bibinfo {author} {\bibfnamefont {G.}~\bibnamefont
			{Aleksandrowicz}}, \bibinfo {author} {\bibfnamefont {T.}~\bibnamefont
			{Alexander}}, \bibinfo {author} {\bibfnamefont {G.}~\bibnamefont
			{Alexandrowics}}, \bibinfo {author} {\bibfnamefont {E.}~\bibnamefont
			{Arbel}}, \bibinfo {author} {\bibfnamefont {A.}~\bibnamefont {Asfaw}},
		\bibinfo {author} {\bibfnamefont {C.}~\bibnamefont {Azaustre}}, \bibinfo
		{author} {\bibfnamefont {P.}~\bibnamefont {Barkoutsos}}, \bibinfo {author}
		{\bibfnamefont {G.}~\bibnamefont {Barron}}, \bibinfo {author} {\bibfnamefont
			{L.}~\bibnamefont {Bello}}, \bibinfo {author} {\bibfnamefont
			{Y.}~\bibnamefont {Ben-Haim}}, \bibinfo {author} {\bibfnamefont {L.~S.}\
			\bibnamefont {Bishop}}, \bibinfo {author} {\bibfnamefont {S.}~\bibnamefont
			{Bosch}}, \bibinfo {author} {\bibfnamefont {D.}~\bibnamefont {Bucher}},
		\bibinfo {author} {\bibnamefont {CZ}}, \bibinfo {author} {\bibfnamefont
			{F.}~\bibnamefont {Cabrera}}, \bibinfo {author} {\bibfnamefont
			{P.}~\bibnamefont {Calpin}}, \bibinfo {author} {\bibfnamefont
			{L.}~\bibnamefont {Capelluto}}, \bibinfo {author} {\bibfnamefont
			{J.}~\bibnamefont {Carballo}}, \bibinfo {author} {\bibfnamefont {C.-F.}\
			\bibnamefont {Chen}}, \bibinfo {author} {\bibfnamefont {A.}~\bibnamefont
			{Chen}}, \bibinfo {author} {\bibfnamefont {R.}~\bibnamefont {Chen}}, \bibinfo
		{author} {\bibfnamefont {J.~M.}\ \bibnamefont {Chow}}, \bibinfo {author}
		{\bibfnamefont {C.}~\bibnamefont {Claus}}, \bibinfo {author} {\bibfnamefont
			{A.~W.}\ \bibnamefont {Cross}}, \bibinfo {author} {\bibfnamefont {A.~J.}\
			\bibnamefont {Cross}}, \bibinfo {author} {\bibfnamefont {J.}~\bibnamefont
			{Cruz-Benito}}, \bibinfo {author} {\bibnamefont {Cryoris}}, \bibinfo {author}
		{\bibfnamefont {C.}~\bibnamefont {Culver}}, \bibinfo {author} {\bibfnamefont
			{A.~D.}\ \bibnamefont {C{\'o}rcoles-Gonzales}}, \bibinfo {author}
		{\bibfnamefont {S.}~\bibnamefont {Dague}}, \bibinfo {author} {\bibfnamefont
			{M.}~\bibnamefont {Dartiailh}}, \bibinfo {author} {\bibfnamefont {A.~R.}\
			\bibnamefont {Davila}}, \bibinfo {author} {\bibfnamefont {D.}~\bibnamefont
			{Ding}}, \bibinfo {author} {\bibfnamefont {E.}~\bibnamefont {Dumitrescu}},
		\bibinfo {author} {\bibfnamefont {K.}~\bibnamefont {Dumon}}, \bibinfo
		{author} {\bibfnamefont {I.}~\bibnamefont {Duran}}, \bibinfo {author}
		{\bibfnamefont {P.}~\bibnamefont {Eendebak}}, \bibinfo {author}
		{\bibfnamefont {D.}~\bibnamefont {Egger}}, \bibinfo {author} {\bibfnamefont
			{M.}~\bibnamefont {Everitt}}, \bibinfo {author} {\bibfnamefont {P.~M.}\
			\bibnamefont {Fern{\'a}ndez}}, \bibinfo {author} {\bibfnamefont
			{A.}~\bibnamefont {Frisch}}, \bibinfo {author} {\bibfnamefont
			{A.}~\bibnamefont {Fuhrer}}, \bibinfo {author} {\bibfnamefont
			{J.}~\bibnamefont {Gacon}}, \bibinfo {author} {\bibnamefont {Gadi}}, \bibinfo
		{author} {\bibfnamefont {B.~G.}\ \bibnamefont {Gago}}, \bibinfo {author}
		{\bibfnamefont {J.~M.}\ \bibnamefont {Gambetta}}, \bibinfo {author}
		{\bibfnamefont {L.}~\bibnamefont {Garcia}}, \bibinfo {author} {\bibfnamefont
			{S.}~\bibnamefont {Garion}}, \bibinfo {author} {\bibnamefont {Gawel-Kus}},
		\bibinfo {author} {\bibfnamefont {L.}~\bibnamefont {Gil}}, \bibinfo {author}
		{\bibfnamefont {J.}~\bibnamefont {Gomez-Mosquera}}, \bibinfo {author}
		{\bibfnamefont {S.}~\bibnamefont {de~la Puente~Gonz{\'a}lez}}, \bibinfo
		{author} {\bibfnamefont {D.}~\bibnamefont {Greenberg}}, \bibinfo {author}
		{\bibfnamefont {J.~A.}\ \bibnamefont {Gunnels}}, \bibinfo {author}
		{\bibfnamefont {I.}~\bibnamefont {Haide}}, \bibinfo {author} {\bibfnamefont
			{I.}~\bibnamefont {Hamamura}}, \bibinfo {author} {\bibfnamefont
			{V.}~\bibnamefont {Havlicek}}, \bibinfo {author} {\bibfnamefont
			{J.}~\bibnamefont {Hellmers}}, \bibinfo {author} {\bibfnamefont
			{{\L}.}~\bibnamefont {Herok}}, \bibinfo {author} {\bibfnamefont
			{H.}~\bibnamefont {Horii}}, \bibinfo {author} {\bibfnamefont
			{C.}~\bibnamefont {Howington}}, \bibinfo {author} {\bibfnamefont
			{W.}~\bibnamefont {Hu}}, \bibinfo {author} {\bibfnamefont {S.}~\bibnamefont
			{Hu}}, \bibinfo {author} {\bibfnamefont {H.}~\bibnamefont {Imai}}, \bibinfo
		{author} {\bibfnamefont {T.}~\bibnamefont {Imamichi}}, \bibinfo {author}
		{\bibfnamefont {R.}~\bibnamefont {Iten}}, \bibinfo {author} {\bibfnamefont
			{T.}~\bibnamefont {Itoko}}, \bibinfo {author} {\bibfnamefont
			{A.}~\bibnamefont {Javadi-Abhari}}, \bibinfo {author} {\bibnamefont
			{Jessica}}, \bibinfo {author} {\bibfnamefont {K.}~\bibnamefont {Johns}},
		\bibinfo {author} {\bibfnamefont {N.}~\bibnamefont {Kanazawa}}, \bibinfo
		{author} {\bibfnamefont {A.}~\bibnamefont {Karazeev}}, \bibinfo {author}
		{\bibfnamefont {P.}~\bibnamefont {Kassebaum}}, \bibinfo {author}
		{\bibfnamefont {V.}~\bibnamefont {Krishnan}}, \bibinfo {author}
		{\bibfnamefont {K.}~\bibnamefont {Krsulich}}, \bibinfo {author}
		{\bibfnamefont {G.}~\bibnamefont {Kus}}, \bibinfo {author} {\bibfnamefont
			{R.}~\bibnamefont {LaRose}}, \bibinfo {author} {\bibfnamefont
			{R.}~\bibnamefont {Lambert}}, \bibinfo {author} {\bibfnamefont
			{J.}~\bibnamefont {Latone}}, \bibinfo {author} {\bibfnamefont
			{S.}~\bibnamefont {Lawrence}}, \bibinfo {author} {\bibfnamefont
			{P.}~\bibnamefont {Liu}}, \bibinfo {author} {\bibfnamefont {P.~B.~Z.}\
			\bibnamefont {Mac}}, \bibinfo {author} {\bibfnamefont {Y.}~\bibnamefont
			{Maeng}}, \bibinfo {author} {\bibfnamefont {A.}~\bibnamefont {Malyshev}},
		\bibinfo {author} {\bibfnamefont {J.}~\bibnamefont {Marecek}}, \bibinfo
		{author} {\bibfnamefont {M.}~\bibnamefont {Marques}}, \bibinfo {author}
		{\bibfnamefont {D.}~\bibnamefont {Mathews}}, \bibinfo {author} {\bibfnamefont
			{A.}~\bibnamefont {Matsuo}}, \bibinfo {author} {\bibfnamefont {D.~T.}\
			\bibnamefont {McClure}}, \bibinfo {author} {\bibfnamefont {C.}~\bibnamefont
			{McGarry}}, \bibinfo {author} {\bibfnamefont {D.}~\bibnamefont {McKay}},
		\bibinfo {author} {\bibfnamefont {S.}~\bibnamefont {Meesala}}, \bibinfo
		{author} {\bibfnamefont {A.}~\bibnamefont {Mezzacapo}}, \bibinfo {author}
		{\bibfnamefont {R.}~\bibnamefont {Midha}}, \bibinfo {author} {\bibfnamefont
			{Z.}~\bibnamefont {Minev}}, \bibinfo {author} {\bibfnamefont
			{R.}~\bibnamefont {Morales}}, \bibinfo {author} {\bibfnamefont
			{P.}~\bibnamefont {Murali}}, \bibinfo {author} {\bibfnamefont
			{J.}~\bibnamefont {M{\"u}ggenburg}}, \bibinfo {author} {\bibfnamefont
			{D.}~\bibnamefont {Nadlinger}}, \bibinfo {author} {\bibfnamefont
			{G.}~\bibnamefont {Nannicini}}, \bibinfo {author} {\bibfnamefont
			{P.}~\bibnamefont {Nation}}, \bibinfo {author} {\bibfnamefont
			{Y.}~\bibnamefont {Naveh}}, \bibinfo {author} {\bibnamefont
			{Nick-Singstock}}, \bibinfo {author} {\bibfnamefont {P.}~\bibnamefont
			{Niroula}}, \bibinfo {author} {\bibfnamefont {H.}~\bibnamefont {Norlen}},
		\bibinfo {author} {\bibfnamefont {L.~J.}\ \bibnamefont {O'Riordan}}, \bibinfo
		{author} {\bibfnamefont {P.}~\bibnamefont {Ollitrault}}, \bibinfo {author}
		{\bibfnamefont {S.}~\bibnamefont {Oud}}, \bibinfo {author} {\bibfnamefont
			{D.}~\bibnamefont {Padilha}}, \bibinfo {author} {\bibfnamefont
			{H.}~\bibnamefont {Paik}}, \bibinfo {author} {\bibfnamefont {S.}~\bibnamefont
			{Perriello}}, \bibinfo {author} {\bibfnamefont {A.}~\bibnamefont {Phan}},
		\bibinfo {author} {\bibfnamefont {M.}~\bibnamefont {Pistoia}}, \bibinfo
		{author} {\bibfnamefont {A.}~\bibnamefont {Pozas-iKerstjens}}, \bibinfo
		{author} {\bibfnamefont {V.}~\bibnamefont {Prutyanov}}, \bibinfo {author}
		{\bibfnamefont {J.}~\bibnamefont {P{\'e}rez}}, \bibinfo {author}
		{\bibnamefont {Quintiii}}, \bibinfo {author} {\bibfnamefont {R.}~\bibnamefont
			{Raymond}}, \bibinfo {author} {\bibfnamefont {R.~M.-C.}\ \bibnamefont
			{Redondo}}, \bibinfo {author} {\bibfnamefont {M.}~\bibnamefont {Reuter}},
		\bibinfo {author} {\bibfnamefont {D.~M.}\ \bibnamefont {Rodr{\'\i}guez}},
		\bibinfo {author} {\bibfnamefont {M.}~\bibnamefont {Ryu}}, \bibinfo {author}
		{\bibfnamefont {M.}~\bibnamefont {Sandberg}}, \bibinfo {author}
		{\bibfnamefont {N.}~\bibnamefont {Sathaye}}, \bibinfo {author} {\bibfnamefont
			{B.}~\bibnamefont {Schmitt}}, \bibinfo {author} {\bibfnamefont
			{C.}~\bibnamefont {Schnabel}}, \bibinfo {author} {\bibfnamefont {T.~L.}\
			\bibnamefont {Scholten}}, \bibinfo {author} {\bibfnamefont {E.}~\bibnamefont
			{Schoute}}, \bibinfo {author} {\bibfnamefont {I.~F.}\ \bibnamefont
			{Sertage}}, \bibinfo {author} {\bibfnamefont {Y.}~\bibnamefont {Shi}},
		\bibinfo {author} {\bibfnamefont {A.}~\bibnamefont {Silva}}, \bibinfo
		{author} {\bibfnamefont {Y.}~\bibnamefont {Siraichi}}, \bibinfo {author}
		{\bibfnamefont {S.}~\bibnamefont {Sivarajah}}, \bibinfo {author}
		{\bibfnamefont {J.~A.}\ \bibnamefont {Smolin}}, \bibinfo {author}
		{\bibfnamefont {M.}~\bibnamefont {Soeken}}, \bibinfo {author} {\bibfnamefont
			{D.}~\bibnamefont {Steenken}}, \bibinfo {author} {\bibfnamefont
			{M.}~\bibnamefont {Stypulkoski}}, \bibinfo {author} {\bibfnamefont
			{H.}~\bibnamefont {Takahashi}}, \bibinfo {author} {\bibfnamefont
			{C.}~\bibnamefont {Taylor}}, \bibinfo {author} {\bibfnamefont
			{P.}~\bibnamefont {Taylour}}, \bibinfo {author} {\bibfnamefont
			{S.}~\bibnamefont {Thomas}}, \bibinfo {author} {\bibfnamefont
			{M.}~\bibnamefont {Tillet}}, \bibinfo {author} {\bibfnamefont
			{M.}~\bibnamefont {Tod}}, \bibinfo {author} {\bibfnamefont {E.}~\bibnamefont
			{de~la Torre}}, \bibinfo {author} {\bibfnamefont {K.}~\bibnamefont
			{Trabing}}, \bibinfo {author} {\bibfnamefont {M.}~\bibnamefont {Treinish}},
		\bibinfo {author} {\bibnamefont {TrishaPe}}, \bibinfo {author} {\bibfnamefont
			{W.}~\bibnamefont {Turner}}, \bibinfo {author} {\bibfnamefont
			{Y.}~\bibnamefont {Vaknin}}, \bibinfo {author} {\bibfnamefont {C.~R.}\
			\bibnamefont {Valcarce}}, \bibinfo {author} {\bibfnamefont {F.}~\bibnamefont
			{Varchon}}, \bibinfo {author} {\bibfnamefont {D.}~\bibnamefont {Vogt-Lee}},
		\bibinfo {author} {\bibfnamefont {C.}~\bibnamefont {Vuillot}}, \bibinfo
		{author} {\bibfnamefont {J.}~\bibnamefont {Weaver}}, \bibinfo {author}
		{\bibfnamefont {R.}~\bibnamefont {Wieczorek}}, \bibinfo {author}
		{\bibfnamefont {J.~A.}\ \bibnamefont {Wildstrom}}, \bibinfo {author}
		{\bibfnamefont {R.}~\bibnamefont {Wille}}, \bibinfo {author} {\bibfnamefont
			{E.}~\bibnamefont {Winston}}, \bibinfo {author} {\bibfnamefont {J.~J.}\
			\bibnamefont {Woehr}}, \bibinfo {author} {\bibfnamefont {S.}~\bibnamefont
			{Woerner}}, \bibinfo {author} {\bibfnamefont {R.}~\bibnamefont {Woo}},
		\bibinfo {author} {\bibfnamefont {C.~J.}\ \bibnamefont {Wood}}, \bibinfo
		{author} {\bibfnamefont {R.}~\bibnamefont {Wood}}, \bibinfo {author}
		{\bibfnamefont {S.}~\bibnamefont {Wood}}, \bibinfo {author} {\bibfnamefont
			{J.}~\bibnamefont {Wootton}}, \bibinfo {author} {\bibfnamefont
			{D.}~\bibnamefont {Yeralin}}, \bibinfo {author} {\bibfnamefont
			{J.}~\bibnamefont {Yu}}, \bibinfo {author} {\bibfnamefont {L.}~\bibnamefont
			{Zdanski}}, \bibinfo {author} {\bibnamefont {Zoufalc}}, \bibinfo {author}
		{\bibnamefont {anedumla}}, \bibinfo {author} {\bibnamefont {azulehner}},
		\bibinfo {author} {\bibnamefont {bcamorrison}}, \bibinfo {author}
		{\bibnamefont {drholmie}}, \bibinfo {author} {\bibnamefont {fanizzamarco}},
		\bibinfo {author} {\bibnamefont {kanejess}}, \bibinfo {author} {\bibnamefont
			{klinvill}}, \bibinfo {author} {\bibnamefont {merav aharoni}}, \bibinfo
		{author} {\bibnamefont {ordmoj}}, \bibinfo {author} {\bibnamefont
			{tigerjack}}, \bibinfo {author} {\bibnamefont {yang.luh}}, \ and\ \bibinfo
		{author} {\bibnamefont {yotamvakninibm}},\ }\href {\doibase
		10.5281/zenodo.2562110} {\enquote {\bibinfo {title} {Qiskit: An open-source
				framework for quantum computing},}\ } (\bibinfo {year} {2019})\BibitemShut
	{NoStop}%
	\bibitem [{\citenamefont {Hill}\ and\ \citenamefont
		{Wootters}(1997)}]{hill1997entanglement}%
	\BibitemOpen
	\bibfield  {author} {\bibinfo {author} {\bibfnamefont {S.}~\bibnamefont
			{Hill}}\ and\ \bibinfo {author} {\bibfnamefont {W.~K.}\ \bibnamefont
			{Wootters}},\ }\href@noop {} {\bibfield  {journal} {\bibinfo  {journal}
			{Physical review letters}\ }\textbf {\bibinfo {volume} {78}},\ \bibinfo
		{pages} {5022} (\bibinfo {year} {1997})}\BibitemShut {NoStop}%
	\bibitem [{\citenamefont {Brunner}\ \emph {et~al.}(2014)\citenamefont
		{Brunner}, \citenamefont {Cavalcanti}, \citenamefont {Pironio}, \citenamefont
		{Scarani},\ and\ \citenamefont {Wehner}}]{brunner2014bell}%
	\BibitemOpen
	\bibfield  {author} {\bibinfo {author} {\bibfnamefont {N.}~\bibnamefont
			{Brunner}}, \bibinfo {author} {\bibfnamefont {D.}~\bibnamefont {Cavalcanti}},
		\bibinfo {author} {\bibfnamefont {S.}~\bibnamefont {Pironio}}, \bibinfo
		{author} {\bibfnamefont {V.}~\bibnamefont {Scarani}}, \ and\ \bibinfo
		{author} {\bibfnamefont {S.}~\bibnamefont {Wehner}},\ }\href@noop {}
	{\bibfield  {journal} {\bibinfo  {journal} {Reviews of Modern Physics}\
		}\textbf {\bibinfo {volume} {86}},\ \bibinfo {pages} {419} (\bibinfo {year}
		{2014})}\BibitemShut {NoStop}%
	\bibitem [{\citenamefont {Werner}(1989)}]{WernerOriginalPaper}%
	\BibitemOpen
	\bibfield  {author} {\bibinfo {author} {\bibfnamefont {R.~F.}\ \bibnamefont
			{Werner}},\ }\href {\doibase 10.1103/PhysRevA.40.4277} {\bibfield  {journal}
		{\bibinfo  {journal} {Phys. Rev. A}\ }\textbf {\bibinfo {volume} {40}},\
		\bibinfo {pages} {4277} (\bibinfo {year} {1989})}\BibitemShut {NoStop}%
	\bibitem [{\citenamefont {V\'ertesi}(2008)}]{WernerFidelityCHSH}%
	\BibitemOpen
	\bibfield  {author} {\bibinfo {author} {\bibfnamefont {T.}~\bibnamefont
			{V\'ertesi}},\ }\href {\doibase 10.1103/PhysRevA.78.032112} {\bibfield
		{journal} {\bibinfo  {journal} {Phys. Rev. A}\ }\textbf {\bibinfo {volume}
			{78}},\ \bibinfo {pages} {032112} (\bibinfo {year} {2008})}\BibitemShut
	{NoStop}%
	\bibitem [{\citenamefont {Lu}\ \emph {et~al.}(2019)\citenamefont {Lu},
		\citenamefont {Li}, \citenamefont {Yin}, \citenamefont {Zhang}, \citenamefont
		{Fang}, \citenamefont {Li}, \citenamefont {Liu}, \citenamefont {Xu},
		\citenamefont {Chen},\ and\ \citenamefont {Pan}}]{lu2019experimental}%
	\BibitemOpen
	\bibfield  {author} {\bibinfo {author} {\bibfnamefont {H.}~\bibnamefont
			{Lu}}, \bibinfo {author} {\bibfnamefont {Z.-D.}\ \bibnamefont {Li}}, \bibinfo
		{author} {\bibfnamefont {X.-F.}\ \bibnamefont {Yin}}, \bibinfo {author}
		{\bibfnamefont {R.}~\bibnamefont {Zhang}}, \bibinfo {author} {\bibfnamefont
			{X.-X.}\ \bibnamefont {Fang}}, \bibinfo {author} {\bibfnamefont
			{L.}~\bibnamefont {Li}}, \bibinfo {author} {\bibfnamefont {N.-L.}\
			\bibnamefont {Liu}}, \bibinfo {author} {\bibfnamefont {F.}~\bibnamefont
			{Xu}}, \bibinfo {author} {\bibfnamefont {Y.-A.}\ \bibnamefont {Chen}}, \ and\
		\bibinfo {author} {\bibfnamefont {J.-W.}\ \bibnamefont {Pan}},\ }\href
	{\doibase 10.1038/s41534-019-0207-2} {\bibfield  {journal} {\bibinfo
			{journal} {npj Quantum Inf}\ }\textbf {\bibinfo {volume} {5}},\ \bibinfo
		{pages} {89} (\bibinfo {year} {2019})}\BibitemShut {NoStop}%
	\bibitem [{\citenamefont {Hayashi}(2007)}]{hayashi2007prior}%
	\BibitemOpen
	\bibfield  {author} {\bibinfo {author} {\bibfnamefont {M.}~\bibnamefont
			{Hayashi}},\ }\href@noop {} {\bibfield  {journal} {\bibinfo  {journal}
			{Physical Review A}\ }\textbf {\bibinfo {volume} {76}},\ \bibinfo {pages}
		{040301} (\bibinfo {year} {2007})}\BibitemShut {NoStop}%
\end{thebibliography}
%

\end{document}